\newif\ifarxiv      
\arxivtrue          

\documentclass[journal,twoside,web]{ieeecolor}
\usepackage{generic}
\usepackage{cite}
\usepackage{amsmath,amssymb,amsfonts}
\usepackage{algorithmic}
\usepackage{graphicx}
\usepackage{algorithm,algorithmic}
\usepackage{hyperref}
\hypersetup{hidelinks}
\usepackage{textcomp}
\usepackage{booktabs}

\usepackage{mathtools}
\usepackage{cases}
\usepackage{soul} 
\usepackage{empheq} 
\usepackage{tcolorbox}
\usepackage[table]{xcolor}
\usepackage{etoolbox}
\usepackage{caption}

\DeclareMathOperator*{\E}{\mathbb{E}}

\newtheorem{lemma}{Lemma}
\newtheorem{proposition}{Proposition}
\newtheorem{theorem}{Theorem}
\newtheorem{corollary}{Corollary}
\newtheorem{remark}{Remark}
\newtheorem{assumption}{Assumption}
\newtheorem{definition}{Definition}

\newcommand{\qed}{\unskip\nobreak\hfill $\square$}
\let\oldtheorem\theorem
\let\oldendtheorem\endtheorem
\def\theorem{\begingroup \oldtheorem}
\def\endtheorem{\qed \oldendtheorem \endgroup}
\let\oldlemma\lemma
\let\oldendlemma\endlemma
\def\lemma{\begingroup \oldlemma}
\def\endlemma{\qed \oldendlemma \endgroup}
\let\oldproposition\proposition
\let\oldendproposition\endproposition
\def\proposition{\begingroup \oldproposition}
\def\endproposition{\qed \oldendproposition \endgroup}
\let\oldcorollary\corollary
\let\oldendcorollary\endcorollary
\def\corollary{\begingroup \oldcorollary}
\def\endcorollary{\qed \oldendcorollary \endgroup}
\let\oldremark\remark
\let\oldendremark\endremark
\def\remark{\begingroup \oldremark}
\def\endremark{\qed \oldendremark \endgroup}
\let\oldassumption\assumption
\let\oldendassumption\endassumption
\def\assumption{\begingroup \oldassumption}
\def\endassumption{\qed \oldendassumption \endgroup}
\let\olddefinition\definition
\let\oldenddefinition\enddefinition
\def\definition{\begingroup \olddefinition}
\def\enddefinition{\qed \oldenddefinition \endgroup}

\makeatletter
\patchcmd{\@makecaption}
  {\scshape}
  {}
  {}
  {}
\patchcmd{\@makecaption}
  {\\}
  {.\ }
  {}
  {}
\makeatother

\newcommand{\aref}[1]{\hyperref[#1]{Appendix~\ref*{#1}}}    

\definecolor{myRed}{HTML}{c44e52}
\definecolor{intractableCol}{HTML}{F19C99}
\definecolor{myBlue}{HTML}{4c72b0}
\definecolor{tractableCol}{HTML}{CCE5FF}
\colorlet{inputsCol}{red!15!blue!15}
\newcommand{\bestres}{green!15}

\newtcolorbox{mytcolorbox}{colback=blue!2!white, colframe=gray!95!black, boxrule=0.1mm, top=0mm, bottom=0mm, left=0mm}
\newtcolorbox{mycolorbox}{colback=blue!2!white, colframe=gray!95!black, boxrule=0mm, top=0mm, bottom=0mm}

\def\BibTeX{{\rm B\kern-.05em{\sc i\kern-.025em b}\kern-.08em
    T\kern-.1667em\lower.7ex\hbox{E}\kern-.125emX}}
\ifarxiv
\else
\fi

\begin{document}
\title{PAC-Bayesian Optimal Control with \\Stability and Generalization Guarantees}

\author{
    Mahrokh Ghoddousi Boroujeni, \IEEEmembership{Student Member, IEEE}, 
    Clara Luc\'{i}a Galimberti, \IEEEmembership{Student Member, IEEE}, 
    Andreas Krause, \IEEEmembership{Senior Member, IEEE},
    and
    Giancarlo Ferrari-Trecate, \IEEEmembership{Senior Member, IEEE}
\thanks{
  This work was supported as a part of NCCR Automation, a National Centre of Competence in Research, funded by the Swiss National Science Foundation (grant number 51NF40\_225155).
  }
\thanks{M. G. Boroujeni and G. Ferrari-Trecate are with the Institute of Mechanical Engineering, EPFL, Switzerland (emails:
\{mahrokh.ghoddousiboroujeni, giancarlo.ferraritrecate\}@epfl.ch).}
\thanks{C. L. Galimberti is with the Dalle Molle Institute for Artificial Intelligence (IDSIA), SUPSI, Switzerland (email: clara.galimberti@supsi.ch).}
\thanks{A. Krause is with the Department of Computer Science, ETH Zürich, Switzerland (email: krausea@ethz.ch).
}}

\maketitle

\begin{abstract}
Stochastic Nonlinear Optimal Control (SNOC) seeks to minimize a cost function that accounts for random disturbances acting on a nonlinear dynamical system. Since the expectation over all disturbances is generally intractable, a common surrogate is the empirical cost, obtained by averaging over a finite dataset of sampled noise realizations. This substitution, however, introduces the challenge of guaranteeing performance under unseen disturbances. The issue is particularly severe when the dataset is limited, as the trained controllers may overfit, leading to substantial gaps between their empirical cost and the deployment cost. In this work, we develop a PAC-Bayesian framework that establishes rigorous generalization bounds for SNOC. Building on these bounds, we propose a principled controller design method that balances empirical performance and prior knowledge. To ensure tractability, we derive computationally efficient relaxations of the bounds and employ approximate inference methods.
Our framework further leverages expressive neural controller parameterizations, guaranteeing closed-loop stability.
Through simulated examples, we highlight how prior knowledge can be incorporated into control design and how more reliable controllers can be synthesized for cooperative robotics.
\end{abstract}


\newcommand{\R}{\mathbb{R}}
\newcommand{\N}{\mathbb{N}}
\newcommand{\Y}{\mathbb{Y}}     

\newcommand{\D}{\mathcal{D}}
\renewcommand{\P}{\mathcal{P}}
\newcommand{\Q}{\mathcal{Q}}

\newcommand{\Emme}[0]{\mathcal{M}}
\newcommand{\rollout}{r}
\newcommand{\KL}{\mathrm{KL}}

\newcommand{\state}{x}
\newcommand{\stateSpace}{\mathbb X}
\newcommand{\stateDim}{{N_x}}
\newcommand{\stateInitNom}{\Bar{x}}

\newcommand{\inp}{u}
\newcommand{\inpSpace}{\mathbb U}
\newcommand{\inpDim}{{N_u}}

\newcommand{\cont}{K}
\newcommand{\contParam}{\theta}
\newcommand{\contParamSet}{\Theta}          
\newcommand{\contParamDim}{{N_\theta}}

\newcommand{\dyn}{f}

\newcommand{\noise}{w}
\newcommand{\noiseSpace}{\mathbb W}

\newcommand{\costTraj}{L}
\newcommand{\costTrue}{\mathcal{L}}
\newcommand{\costEmp}{\hat{\mathcal{L}}}
\newcommand{\costBound}{C}
\newcommand{\costDen}{\gamma}

\renewcommand{\S}{\mathbb{S}}
\newcommand{\numRollouts}{S}
\newcommand{\counterRollout}{s}

\newcommand{\confidence}{\delta}
\newcommand{\tempGibbs}{\lambda}
\newcommand{\tempGibbsGrowth}{d}

\newcommand{\particle}{\phi}
\newcommand{\particles}{\Phi}
\newcommand{\numParticles}{k}
\newcommand{\counterParticle}{\kappa}

\newcommand{\numPSamples}{N_P}
\newcommand{\counterPSamples}{n_P}

\newcommand{\numQSamples}{N_Q}
\newcommand{\counterQSamples}{n_Q}

\newcommand{\timestep}{t}
\newcommand{\horizon}{T}

\newcommand{\lp}{\ell_p}

\section{Introduction}
\label{sec:introduction}
\IEEEPARstart{S}{tochastic} Nonlinear Optimal Control (SNOC) concerns minimizing a cost function that accounts for stochastic uncertainties affecting nonlinear systems, such as process noise and disturbances. 
Beyond the linear quadratic Gaussian setting, this problem is generally intractable. It is often replaced by minimizing the average cost over a dataset of sampled uncertainties, i.e., the empirical cost. This substitution raises the fundamental challenge of deriving generalization bounds, namely, upper bounds on the true (intractable) cost in terms of its empirical (computable) counterpart.

A closely related problem has been extensively studied in supervised machine learning within the Probably Approximately Correct (PAC)-Bayesian framework~\cite{userfriendly}. PAC-Bayesian analysis provides high-probability generalization bounds for predictors sampled from an arbitrary posterior distribution, either in expectation or for a single draw. The calculation of these bounds typically comprises the empirical cost and a discrepancy measure between the posterior and a prior distribution.

Classic PAC-Bayesian bounds for supervised learning require \emph{aggregated predictors}, obtained by averaging all predictors according to the posterior. Transferring this idea to SNOC would imply aggregating control actions across all possible controllers weighted by the posterior, which is computationally infeasible. 
Instead, we adopt a less commonly used variant of PAC-Bayesian bounds that applies to \emph{randomized predictors}~\cite{userfriendly}, i.e., single predictors drawn from the posterior. In the SNOC context, this enables the establishment of generalization guarantees for a \emph{single controller} sampled from the posterior, which is far more practical in control applications. The proposed PAC-Bayesian bound complements existing approaches to certifying learned controllers, such as those relying on mixed-integer programming or satisfiability modulo theory techniques~\cite{dawson2023safe, schwan2023stability}.
A discussion of related work on PAC-Bayesian control can be found in \aref{app:relatedworkPAC}.

Beyond generalization, the sampled controller must achieve low cost and guarantee closed-loop stability, which is essential in control theory. This can be accomplished by defining prior and posterior distributions over families of controllers that are expressive enough to achieve high performance on complex tasks, yet restricted to ensure stability. To this end, we leverage the Performance-Boosting framework~\cite{NeurSLS,furieri2024learning}, which provides broad classes of nonlinear neural network (NN)-based controllers with closed-loop stability guarantees. It further enables the optimization of general cost functions, which can reflect diverse closed-loop specifications, from minimizing the L$_2$-gain of the closed-loop system to encoding safety. Importantly, it provides an unconstrained parametrization of all and only stabilizing controllers (in the $\lp$ sense) for a given $\lp$-stable nonlinear system. Since this framework guarantees closed-loop stability for any parameter choice, any posterior distribution over this parameter space assigns probability mass exclusively to stabilizing controllers.

\subsection{Contributions}
We propose a PAC-Bayesian framework for SNOC that advances both the theoretical foundations and the algorithmic methodology of control design. 
On the theoretical side, we \textit{(\romannumeral 1)} establish generalization certificates for SNOC and \textit{(\romannumeral 2)} derive a computationally efficient, albeit looser, variant that constitutes the first tractable PAC-Bayesian bound for SNOC, enabling scalability to realistic control problems. 
On the algorithmic side, we \textit{(\romannumeral 3)} develop a control policy learning procedure that allows sampling controllers from an optimal posterior distribution constructed to minimize the generalization bound in expectation, \textit{(\romannumeral 4)} implement highly expressive neural controllers with built-in stability guarantees~\cite{furieri2024learning}, and \textit{(\romannumeral 5)} approximate the optimal posterior to make controller sampling numerically tractable. 
We validate our framework in two examples. First, we show how coarse prior knowledge can be incorporated into control design using a toy example. Second, we demonstrate its potential in a cooperative robotic task.

This paper builds upon our conference version~\cite{Boroujeni2024PACSNOC}, where we introduced the first PAC-Bayesian approach for SNOC. The present work extends those results by deriving both upper and lower bounds on the true cost, introducing novel tractable bounds, and broadening the controller architectures to include State-Space Models (SSMs)~\cite{gu2022efficiently} alongside Recurrent Equilibrium Networks (RENs)~\cite{revay2023recurrent}. For approximating the posterior distribution, we incorporate normalizing flows~\cite{rezende2015variational} in addition to Stein Variational Gradient Descent (SVGD)~\cite{liu2016stein}. Furthermore, we propose constructing a data-driven prior distribution that helps in tightening the bounds. Finally, we enhance performance with a bootstrapping strategy. The advantages of these extensions are validated numerically.

\subsection{Notation}
$\N_0$ denotes $\N \cup \{0\}$. $\R^+$ ($\R^+_0$) stands for positive (non-negative) real numbers.
We denote the truncated sequence $(x_0,x_1,\ldots,x_{t})$ as $x_{t:0}$, where $x_{\tau}\in\R^n$ for $\tau \in \{0, \cdots, t\}$. In the limit, as $t\rightarrow \infty$, this sequence is denoted as $x_{\infty:0}\coloneqq (x_0,x_1,\ldots)$.
The set of all sequences $x_{\infty:0}$ is denoted as $\ell^n$. 
Similar to sequences, we define $A_{t:0}(x_{t:0}) = (A_0(x_0), A_1(x_{1:0}), \dots, A_t(x_{t:0}))$ where each $A_\tau$ is a function.
Finally, the zero vector in $\R^n$ is denoted as $0_n$.

\section{Stable SNOC}
We begin with an overview of closed-loop stability and SNOC to establish the basis for subsequent analysis.

\subsection{Stable closed-loop systems}

Consider a discrete-time system with state $\state_t \in \stateSpace \subset \mathbb{R}^{\stateDim}$ and input $\inp_t \in \inpSpace \subset \mathbb{R}^{\inpDim}$ at time $t \in \mathbb{N}_0$. The initial state $\state_0$ has a nominal value $\stateInitNom \in \stateSpace$ and evolves according to:
\begin{subequations}
\label{eq:system}
\begin{align}
    \state_0 &= \stateInitNom + \noise_0, \label{eq:system_ini} \\
    \state_t &= \dyn_t (\state_{t-1:0}, \inp_{t-1:0}) + \noise_t, \quad t =1,2,\dots. \label{eq:system_step}
\end{align}
\end{subequations}
The dynamics $\dyn_t : \stateSpace^{t}\times\inpSpace^{t} \rightarrow \stateSpace$, which may be nonlinear and time-varying, is assumed to be known. Process noise $\noise_t \in \noiseSpace \subset \mathbb{R}^{\stateDim}$ is drawn from a distribution $\D_t$, with $\D_0$ specifically modeling initial state uncertainty. While the exact probability density of $\D_t$ is unknown, we assume access to samples from it.%
\footnote{In~\eqref{eq:system}, $\dyn_t$ and $\noise_t$ on the right-hand side are indexed by $t$ rather than $t{-}1$. This is done to simplify the notation by including the uncertainty on $\state_0$ as the first element of the sequence $(\noise_0, \noise_1, \dots)$. An equivalent formulation with indices $t{-}1$ on the right-hand side is given in~\cite{furieri2024learning}.
}

To control the system~\eqref{eq:system}, we employ a time-varying dynamical feedback controller, defined as:
\begin{equation}
    \label{eq:control}
    \inp_t = -\cont_{t}^{\contParam} (\state_{t:0}), \quad \forall t \in \N_0,
\end{equation}
where $\contParam \in \contParamSet \subset \R^{\contParamDim}$ denotes the controller parameters.
Using this controller, the closed-loop system~\eqref{eq:system}-\eqref{eq:control} can be simulated over a horizon $T\in\N$ for a given sequence of noise samples $\noise_{T:0}$, to obtain state and input trajectories, $x_{T:0}$ and $u_{T:0}$.
This relationship is captured by the \emph{rollout} map, $\rollout^{\contParam}_{T:0} : \noiseSpace^{T+1} \rightarrow (\stateSpace {\times} \inpSpace)^{T+1}$, with components $\rollout^{\contParam}_t (\noise_{t:0}) = (\state_t, \inp_t)$ for $t \in \{0, \cdots, T\}$. The rollout map additionally depends on the nominal initial state $\stateInitNom$ and the system dynamics $\dyn$, but we omit these from the notation for simplicity.

Stability is a fundamental property of closed-loop systems, ensuring that bounded noise does not result in uncontrolled growth of the state $\state_t$ or the input $\inp_t$ over time. In this work, we analyze stability within the framework of $\lp$-stability, defined as follows:
\begin{definition}[$\lp$-sequences]\label{def:lp_seq}
    A sequence $z_{\infty:0}$ belongs to $\lp$ for $p \in \mathbb{N} \cup \{\infty\}$ if $\Vert z_{\infty:0} \Vert_p = \bigl( \sum_{t=0}^\infty \vert z_t \vert^p \bigr )^{\frac{1}{p}} < \infty$, where $\vert \cdot \vert$ denotes any vector norm. 
\end{definition}
\begin{definition}[$\lp$-stable mappings]\label{def:lp_op}
    A mapping from a sequence to another sequence is called $\lp$-stable if it transforms every $\lp$ sequence into another $\lp$ sequence.
\end{definition}

Using these definitions, we can formalize $\lp$-stability for closed-loop systems. Recall that a closed-loop system, as per~\eqref{eq:system}–\eqref{eq:control}, can be equivalently described by its rollout map $\rollout^{\contParam}_{\infty:0}$ over all time steps. 
\begin{definition}[$\lp$-stable closed-loop systems]\label{def:lp_closedloop}
    A closed-loop system is called $\lp$-stable if its rollout map $\rollout^{\contParam}_{\infty:0}$ is an $\lp$-stable mapping.
\end{definition}
Intuitively, this means that the system’s energy remains bounded whenever the input noise energy is bounded~\cite{DesoerVidyasagar1975}.

\subsection{Parameterizing stabilizing controllers}
\label{subsec:controller_parametrization}
We now address the construction of the controller $\cont_t^{\contParam}$ to ensure $\lp$-stability of the closed-loop system. To this end, we adopt the \emph{unconstrained parameterization} of stabilizing nonlinear control policies proposed in~\cite{NeurSLS, furieri2024learning}. This parameterization applies to nonlinear time-varying systems, as described in~\eqref{eq:system}, that are $\lp$-stable or have been pre-stabilized.%
\footnote{For readability, we assume that the base controller in~\cite{NeurSLS, furieri2024learning} is incorporated into the system dynamics $\dyn_t$ when necessary.} 
In other words, if $\noise_{\infty:0}$ and $\inp_{\infty:0}$ belong to $\lp$, then the state trajectory $\state_{\infty:0}$ obtained through~\eqref{eq:system} also lies in $\lp$. 
This setting focuses on disturbance sequences with finite $\lp$ norm and therefore excludes unbounded-energy models such as i.i.d. Gaussian noise. In many practical scenarios, however, stochastic disturbances can be represented or approximated by bounded-energy processes that fall within this framework.
Compared to constrained formulations, this unconstrained parameterization scales to controllers with orders of magnitude more parameters, as further discussed in \aref{app:relatedworkPB}.

The framework builds on the internal model control (IMC) paradigm~\cite{Economou_Morari_nlIMC_1986}. Specifically, knowledge of the system dynamics $\dyn$ is used to reconstruct the disturbance $\hat{\noise}_t \in \R^{\stateDim}$, which is then mapped to the control input $\inp_t$. This yields the following state-feedback dynamical controller:
\begin{subequations}
\label{eq:neurSLS}
\begin{empheq}[left= \cont_t^{\contParam} : \empheqlbrace]{align}
  \hat{\noise}_t &= \state_t - \dyn_t(\state_{t-1:0}, \inp_{t-1:0}), 
  \label{eq:neurSLS_w}
  \\
  \inp_t &= \Emme_t^{\contParam}(\hat{\noise}_{t:0}), 
  \label{eq:neurSLS_output}
\end{empheq}
\end{subequations}
where $\Emme_t^{\contParam}$ is an arbitrary function with parameters $\contParam$.%
\footnote{The analysis in~\cite{NeurSLS, furieri2024learning} also extends to non-parametric functions $\Emme_t$, but here we restrict attention to parametric ones for clarity.}

The key result from~\cite{NeurSLS, furieri2024learning} establishes that for an $\lp$-stable system~\eqref{eq:system}, the closed-loop map remains $\lp$-stable if and only if the sequence-to-sequence mapping $\Emme_{\infty:0}^{\contParam}=(\Emme_{0}^{\contParam}, \Emme_{1}^{\contParam}, \ldots)$ is $\lp$-stable. Consequently, designing a stabilizing controller reduces to choosing an $\lp$-stable $\Emme_{\infty:0}^{\contParam}$. This can be achieved using various NN models based on dynamical systems, such as RENs~\cite{revay2023recurrent}, certain classes of SSMs~\cite{gu2022efficiently}, and neural network parameterizations of Hamiltonian systems~\cite{zakwan2024neural}. In our experiments, we employ RENs and SSMs, which are briefly described in \aref{app:implementation}.
Once the architecture of $\Emme_{\infty:0}^{\contParam}$ is fixed, the controller~\eqref{eq:neurSLS_w}–\eqref{eq:neurSLS_output} guarantees stability of the system~\eqref{eq:system} for every parameter choice $\contParam \in \contParamSet = \R^\contParamDim$.%
\footnote{Since the architecture of the controller is fixed, we use $\cont^{\contParam}$ and $\contParam$ interchangeably to refer to the controller and its parameters.}

\subsection{Stable stochastic optimal control}
\label{subsec:snoc}
Having ensured that the controller $\cont_t^{\contParam}$ is stabilizing for any parameter choice, we optimize its parameters $\contParam$ based on a \emph{finite-horizon (FH) cost} $\costTraj(\state_{T:0}, \inp_{T:0})$ over a horizon $T \in \mathbb{N}$. Since trajectories satisfy $(\state_{T:0}, \inp_{T:0}) = \rollout^{\contParam}_{T:0} (\noise_{T:0})$, we equivalently write the FH cost as $\costTraj(\contParam, \noise_{T:0})$. The function $\costTraj: \contParamSet \times \noiseSpace^{T+1} \to \mathbb{R}_+$ is assumed to be nonnegative and piecewise differentiable.

As the noise sequence $\noise_{T:0}$ is stochastic, we define the \emph{true cost} as the expectation of the FH cost under the distribution $\D_{T:0} \coloneqq (\D_0,\ldots,\D_T)$:
\begin{align} 
    \costTrue(\contParam, \D_{T:0})
    &\coloneqq \E_{\noise_{T:0} \sim \D_{T:0}}
        \costTraj(\contParam, \noise_{T:0}). \label{eq:true_cost}
\end{align}
By averaging over noise realizations, $\costTrue$ provides a robust measure of the controller's performance. 

Computing $\costTrue$ in~\eqref{eq:true_cost} is generally intractable as it involves a high-dimensional expectation over $\noise_{T:0}$. For nonlinear systems and policies, there is no closed form, and an accurate Monte Carlo estimation can demand prohibitively many rollouts even when $\D_{T:0}$ is known. Instead, SNOC methods typically optimize the \emph{empirical cost} based on observed data
$\S \coloneqq \bigl\{ \noise_{T:0}^{i} \bigr\}_{\counterRollout=1}^{\numRollouts}$,
where $\numRollouts \in \mathbb{N}$ denotes the number of sampled noise sequences. The empirical cost is defined as:
\begin{align} 
    \costEmp(\contParam, \S) &\coloneqq \frac{1}{\numRollouts} \sum_{\counterRollout=1}^{\numRollouts} \costTraj(\contParam, \noise_{T:0}^\counterRollout), \label{eq:emp_cost} 
\end{align}
which averages the FH cost over the noise sequences in $\S$. Unlike $\costTrue$, $\costEmp$ can be efficiently computed and optimized, leading to the following
empirical SNOC problem:
\begin{mytcolorbox}
    \begin{subequations}\label{eq:SNOC}
    \begin{align}
        \text{Empirical SNOC}:~ \min_{\contParam \in \contParamSet} \; & \costEmp(\contParam, \S), \label{eq:snoc_ob}\\
        \text{s.t.}\; &\rollout^{\contParam}_{\infty:0} \text{ is } \lp\text{-stable}. \label{eq:snoc_stability}
    \end{align}
    \end{subequations}    
\end{mytcolorbox}
\noindent
This formulation minimizes $\costEmp(\contParam, \S)$ while ensuring stability according to \autoref{def:lp_closedloop}. We refer to its solution as the \emph{empirical controller}.


\section{Inference in controller space}\label{sec:PAC}
A critical drawback of the empirical SNOC approach in~\eqref{eq:SNOC} is that the trained controller may perform significantly worse when confronted with out-of-sample noise sequences than those in its training dataset $\S$. Formally, the true cost $\costTrue$ can substantially exceed the empirical cost
$\costEmp$, a phenomenon corresponding to \emph{overfitting} in supervised learning. In this section, we overcome this limitation by establishing a generalization bound on the true cost and using it to propose a controller design algorithm.

We adopt a probabilistic view over controllers: a \emph{prior} distribution $\P$ is specified over the parameter space $\contParamSet$, independently of the dataset $\S$, and a data-dependent \emph{posterior} $\Q$
is formed after observing $\S$. In our setting, both $\P$ and $\Q$ are supported on stabilizing controllers (cf. \autoref{subsec:controller_parametrization}),
so any $\contParam \sim \Q$ is stabilizing by design. 

While the prior and posterior terms resemble Bayesian terminology, they are not necessarily linked by Bayes' law. In classical Bayesian inference, the posterior distribution depends on the observed data exclusively through the likelihood, which measures how probable the data is under a given model~\cite{krause2025prob}. In our context, this would mean evaluating the likelihood of observing a set of noise sequences $\S$ given a controller $\cont^{\contParam}$. However, since the process noise $\noise_t$ is independent of the controller choice, the resulting posterior would be uninformative for control design. Instead, the posterior must reflect how good a controller $\cont^{\contParam}$ is on the dataset $\S$, which we quantify via the empirical cost $\costEmp(\contParam, \S)$. In other words, controllers achieving lower cost on $\S$ should be assigned higher posterior probability relative to their prior. The PAC-Bayesian framework enables precisely this type of data dependence, by constructing posteriors that trade off empirical performance and proximity to the prior.


Later, in \autoref{corol:qstar}, we exploit this flexibility to define the \emph{optimal Gibbs posterior} $\Q^*$, which minimizes the bound in expectation. In practice, we approximate $\Q^*$ using scalable variational methods such as SVGD or normalizing flows, as detailed in \autoref{subsec:approx_Q}.

\subsection{PAC-Bayes generalization bounds for SNOC}\label{subsec:PAC4SNOC}
We begin by bounding the performance gap between true and empirical costs using the PAC-Bayes framework.
Originally developed for supervised learning (e.g.,~\cite{userfriendly}), where the goal is to train predictors with guarantees on unseen data, this approach provides high-probability bounds on generalization performance. In the SNOC setting, we reinterpret these tools by shifting the focus from predictors to controllers trained on datasets of noise sequences. \autoref{tab:sup-snoc} summarizes the correspondence between supervised learning and SNOC. By aligning these correspondences, we effectively extend PAC-Bayesian bounds from supervised learning to the SNOC framework.

\begin{table}[t]
\caption{Correspondence between elements in supervised learning and SNOC.}
\label{tab:sup-snoc}
\setlength{\tabcolsep}{3pt}
\begin{center}
\begin{tabular}{ll|ll}
\toprule
\multicolumn{2}{c|}{Supervised Learning} & \multicolumn{2}{c}{SNOC} 
\\ 
\midrule
Sample & $(x,y) \in \stateSpace {\times} \Y$             & Noise sequence & $\noise_{T:0} \in \noiseSpace^{T+1}$           \\
Predictor & $h^{\contParam}: \stateSpace {\xrightarrow[]{}} \Y$ & Rollout        & $\rollout^{\contParam}_{T:0}: \noiseSpace^{T+1} {\xrightarrow[]{}} (\stateSpace{\times} \inpSpace)^{T+1}$ \\
Loss                & $l(\contParam, x, y)$               & FH cost          &  $L(\contParam, \noise_{T:0})$      \\
\bottomrule
\end{tabular}
\end{center}
\end{table}

Classical PAC-Bayesian bounds deal with the loss of
\emph{aggregated predictors}, i.e., averages with respect to $\Q$. Translated to control, this would require computing control actions by averaging across all sampled controllers, i.e., $\inp_t = - \E_{\contParam \sim \Q}\, \cont^{\contParam}_t(\state_{t:0})$, which is both computationally burdensome and undesirable since a single
controller is typically deployed. We therefore use a less common family of PAC-Bayes bounds that apply to \emph{randomized predictors}, where at deployment, a single controller $\contParam \sim \Q$ is sampled and
used to generate $\inp_t = -\cont^{\contParam}_t(\state_{t:0})$.
This approach aligns well with standard control system practices as it considers a single controller rather than relying on aggregation.

To facilitate the analysis, we impose the following assumption:
\begin{assumption}\label{assumption}
    The FH cost is upper-bounded by $\costBound \in \R^+$ over
    $\contParam \in \contParamSet$ and $\noise_{T:0} \in \noiseSpace^{T+1}$,
    i.e., $\costTraj(\contParam, \noise_{T:0}) \in [0,\costBound)$.
\end{assumption}

\autoref{assumption} may not apply to many cost functions.
For example, even a simple quadratic cost can become unbounded if the closed-loop system is unstable. To address this, we propose mapping any unbounded cost function to $[0, \costBound)$ using:
\begin{align}
    \Tilde{\costTraj}(\contParam, \noise_{T:0}) \coloneqq \costBound \, \tanh \bigr( \costTraj(\contParam, \noise_{T:0}) / \costDen \bigl), \label{eq:cost_trans}
\end{align}
where $\tanh$ is the hyperbolic tangent function and $\costDen \in \mathbb{R}^+$ is a constant. This transformed cost grows almost linearly with the original cost between $0$ and $\costDen$, then smoothly saturates at the maximum value of $\costBound$. A practical choice for $\costDen$ could be the open-loop noise-free cost, $\costDen = \costTraj(\cdot, 0_T)$. Since we generally expect the closed-loop to perform better than the open-loop, this choice keeps the transformed cost predominantly within the linear region rather than near saturation.
In the sequel, we work with the transformed cost $\Tilde{\costTraj}$, which satisfies \autoref{assumption}, and denote it again by $\costTraj$, with a slight abuse of notation.

We now state the randomized PAC-Bayes bounds specialized to SNOC. This theorem and its proof are adapted from
Theorem~2.7 in~\cite{userfriendly} using the relations outlined in \autoref{tab:sup-snoc}.
\begin{theorem}
\label{theo:pac} 
    Let $\S$ be a dataset consisting of $\numRollouts$ noise sequences sampled from $\D_{T:0}$.
    Fix a prior $\P$ independent of $\S$ and any posterior $\Q$.
    Then, for any $\tempGibbs > 0$, confidence level $\confidence \in (0,1)$, and controller parameters $\contParam \sim \Q$, 
    \begin{align}
        \costTrue(\contParam, \D_{T:0})
        &\leq   
        \costEmp(\contParam, \S)
        +
        \frac{1}{\tempGibbs} \ln \frac{d\Q}{d\P}(\contParam) 
        + 
        \frac{1}{\tempGibbs} \ln \frac{1}{\confidence}
        + 
        \frac{\tempGibbs \costBound^2}{8 \numRollouts}, 
        \label{eq:u_bound} \\
        \costTrue(\contParam, \D_{T:0})
        &\geq   
        \costEmp(\contParam, \S)
        -
        \frac{1}{\tempGibbs} \ln \frac{d\Q}{d\P}(\contParam) 
        - 
        \frac{1}{\tempGibbs} \ln \frac{1}{\confidence}
        - 
        \frac{\tempGibbs \costBound^2}{8 \numRollouts}.
        \label{eq:l_bound}
    \end{align}
    Each inequality holds with probability at least $1-\confidence$, and both hold simultaneously with probability at least $1-2\confidence$, where the probability is over jointly sampling $\S \sim \D_{T:0}^{\numRollouts}$ and $\contParam \sim \Q$. The costs $\costTrue$ and $\costEmp$ are as in \eqref{eq:true_cost} and \eqref{eq:emp_cost} and rely on an FH cost satisfying \autoref{assumption}.
\end{theorem}

The inequalities in
\autoref{theo:pac} bound the true cost of any sampled controller $\contParam \sim \Q$ via its empirical cost, a posterior–prior discrepancy term $\ln \tfrac{d\Q}{d\P}(\contParam)$, and constants. Combining \eqref{eq:u_bound}–\eqref{eq:l_bound} yields a bound on the generalization gap
$\bigl|\costTrue(\contParam,\D_{T:0}) - \costEmp(\contParam,\S)\bigr|$, that
shrinks with $\numRollouts$ and vanishes as $\numRollouts \to \infty$ under a suitable scaling of $\tempGibbs$ (e.g., $\tempGibbs \propto
\numRollouts^{1/\tempGibbsGrowth}$ with $\tempGibbsGrowth>1$). 
In \autoref{prop:tempGibbs}, we present a specific choice of $\tempGibbs$ that meets this requirement.

\subsection{Optimal posterior}\label{subsec:qstar}
\subsubsection{Defining the optimal posterior}
As highlighted earlier in this section, $\P$ and $\Q$ are not necessarily linked by  Bayes' law. The flexibility in choosing $\Q$ can be exploited to align the posterior more closely with the SNOC cost $\costTraj$. Ideally, for a given $\contParam$ sampled from $\Q$, one would choose $\Q$ to minimize the upper bound on $\costTrue$ in~\eqref{eq:u_bound}. However, this leads to a circular dependency, since $\Q$ must be fixed before sampling  $\contParam$. To resolve this, we select $\Q$ by minimizing the expected upper bound over draws of the controller parameters, leading to the following SNOC problem:
\begin{mytcolorbox}
    \begin{subequations}\label{eq:PACSNOC}
    \begin{align}
        \text{PAC-SNOC}:~ 
        \min_{\Q}\; & \E_{\contParam \sim \Q} \Bigl[
            \costEmp(\contParam, \S)
            + 
            \frac{1}{\tempGibbs} \ln \frac{d\Q}{d\P}(\contParam)
        \Bigr], \label{eq:pacsnoc_ob}\\
        \text{s.t.}\; &\rollout^{\contParam}_{\infty:0} \text{ is } \lp\text{-stable} \quad \forall \,\contParam \sim \Q, \label{eq:pacsnoc_stability}
    \end{align}
    \end{subequations}    
\end{mytcolorbox}
\noindent
where constant terms in~\eqref{eq:u_bound} have been dropped. The constraint \eqref{eq:pacsnoc_stability} requires that every controller $\contParam$ sampled from the posterior distribution $\Q$ stabilizes the closed-loop system in the sense of \autoref{def:lp_closedloop}.
We refer to the solution of~\eqref{eq:PACSNOC} as the 
\emph{optimal posterior} distribution and denote it by $\Q^*$.

The constraint~\eqref{eq:pacsnoc_stability} imposes closed-loop stability for every $\contParam$ sampled from $\Q$. This is enforced by defining $\Q$ over the stabilizing controller family introduced in \autoref{subsec:controller_parametrization}, which guarantees that any sampled controller satisfies the condition. Under this setup, the optimal posterior is obtained by the unconstrained optimization \eqref{eq:pacsnoc_ob}, ignoring \eqref{eq:pacsnoc_stability}, and admits the following closed-form expression.
\begin{corollary}[Lemma~1.1.3 in~\cite{Catoni}]\label{corol:qstar}
    Given a dataset $\S$, a prior distribution $\P$ chosen independently of $\S$, and a parameter $\tempGibbs > 0$, the solution of \eqref{eq:PACSNOC} is the \textit{Gibbs} distribution:
    \begin{align} 
        \Q^*(\contParam) 
        =& 
            \P(\contParam) \,
            e^{-\tempGibbs \costEmp(\contParam, \S)}
            / Z_\tempGibbs(\P,\S) 
        \label{eq:qstar},\\
        Z_\tempGibbs(\P,\S) 
        \coloneqq& 
            \E_{\contParam \sim \P} 
            e^{-\tempGibbs \costEmp(\contParam, \S)}, 
        \label{eq:Z}
    \end{align}
    where $Z_\tempGibbs(\P,\S)$, known as the \emph{partition 
    function}, normalizes $\Q^*$.   
\end{corollary}

Computing the partition function $Z_\tempGibbs(\P,\S)$ is notoriously difficult, as it requires integration over $\contParam \sim \P$~\cite{holz2023convergence}. This challenge is addressed further in \autoref{subsec:approx_ub}.

\subsubsection{Simplified bounds for the optimal posterior}
Recall that \autoref{theo:pac} provides bounds valid for any posterior distribution $\Q$. Building on \autoref{corol:qstar}, where we established the optimal posterior $\Q^*$, we now tailor \autoref{theo:pac} to this choice.
\begin{corollary}\label{corol:ub_qstar}
    If the optimal posterior $\Q^*$, given by~\eqref{eq:qstar}-\eqref{eq:Z}, is employed in \autoref{theo:pac}, the inequalities,
    \begin{align}
        \costTrue (\contParam, \D_{T:0}) 
        &\leq
        - \frac{1}{\tempGibbs} \ln Z_\tempGibbs(\P,\S)
        + \frac{1}{\tempGibbs} \ln \frac{1}{\confidence}
        + \frac{\tempGibbs \costBound^2}{8 \numRollouts} ,
        \label{eq:u_bound_qstar}
        \\
        \costTrue(\contParam, \D_{T:0})
        &\geq   
        2 \costEmp(\contParam, \S)
        +
        \frac{1}{\tempGibbs} \ln Z_\tempGibbs(\P,\S)
        - 
        \frac{1}{\tempGibbs} \ln \frac{1}{\confidence}
        - 
        \frac{\tempGibbs \costBound^2}{8 \numRollouts}, 
        \label{eq:l_bound_qstar}
    \end{align}
    hold individually with probability at least $1-\confidence$ and jointly with probability at least $1-2\confidence$. The probability is over simultaneously sampling $\S \sim \D_{T:0}^\numRollouts$ and $\contParam \sim \Q^*$.
\end{corollary}

The proof follows by substituting~\eqref{eq:qstar} into~\eqref{eq:u_bound} and~\eqref{eq:l_bound}. 
These bounds, specialized to $\Q^*$, are tighter than the general PAC-Bayes bounds in \autoref{theo:pac}, which hold for any $\Q$, and no longer depend explicitly on the posterior distribution.

\subsubsection{Selecting the Gibbs posterior parameter}
The probability mass assigned to each controller $\contParam$ by the optimal posterior distribution depends on two terms: the empirical cost $\costEmp(\contParam, \S)$ and the prior probability $\P(\contParam)$. The trade-off between these two terms is modulated by the parameter $\tempGibbs$. Equivalently, this can be seen from the objective~\eqref{eq:pacsnoc_ob}, which balances minimizing the expected empirical cost against encouraging similarity between posterior and prior, measured by the KL divergence.

In the extremes, $\Q^*$ collapses to a degenerate distribution at the minimizer(s) of $\costEmp$ as $\tempGibbs \rightarrow \infty$, and converges to the prior $\P$ as $\tempGibbs \rightarrow 0$. The prior can thus be chosen to incorporate existing knowledge or to regularize the posterior complexity, thereby mitigating overfitting (see \autoref{sec:experiments} for examples).

The bounds in \autoref{corol:ub_qstar} become tighter as the right-hand sides decrease, motivating the tuning of the free parameter $\tempGibbs$. However, this is not straightforward since both~\eqref{eq:u_bound_qstar} and~\eqref{eq:l_bound_qstar} depend on the partition function~\eqref{eq:Z}, which is generally intractable. Under \autoref{assumption}, the partition function satisfies $Z_\tempGibbs(\P,\S) \in [e^{-\tempGibbs \costBound}, 1]$, which allows us to derive a tractable but looser relaxation of these bounds. Based on this relaxation, the following result provides a principled choice for $\tempGibbs$.

\begin{proposition}\label{prop:tempGibbs}\looseness=-1
    Consider the relaxation of \eqref{eq:u_bound_qstar}–\eqref{eq:l_bound_qstar} obtained by bounding the partition function.
    The value of $\tempGibbs > 0$ that simultaneously minimizes the relaxed upper bound and maximizes the relaxed lower bound is:
    \begin{align}
        \tempGibbs^* = \sqrt{8 \, \numRollouts \, \ln (1/\confidence)} / C. \label{eq:lambdastar}
    \end{align}
\end{proposition}

Given the prior $\P$ and the dataset $\S$, the Gibbs distribution $\Q^*$ with the parameter $\tempGibbs^*$ defined in \eqref{eq:qstar} and \eqref{eq:lambdastar} fully specifies a posterior distribution, requiring no additional free parameters. This construction provides a principled balance between fitting the data and incorporating prior knowledge.

\begin{figure}[t]
  \centering
  \includegraphics[width=\linewidth]{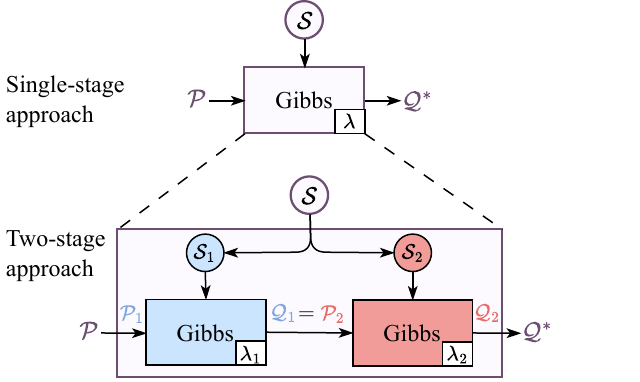}
  \caption{Two-stage inference approach. In the first stage, a subset $\S_1 \subset \S$ together with a data-independent prior $\P_1 = \P$ is used to infer a posterior $\Q_1$. In the second stage, the remaining data $\S_2 = \S \setminus \S_1$ is combined with $\P_2 = \Q_1$ to infer a posterior $\Q_2$. Both stages follow the inference procedure of \autoref{corol:qstar}. When the Gibbs parameters $\tempGibbs_1$ and $\tempGibbs_2$ are chosen according to \autoref{prop:twostage}, the resulting posterior coincides with that of the single-stage procedure, i.e., $\Q_2 = \Q^*$.}
  \label{fig:twostage}
\end{figure}

\subsection{Two-stage inference}\label{subsec:twostage}
The bounds in \autoref{corol:ub_qstar} depend on the prior $\P$ through the partition function $Z_\tempGibbs(\P, \S)$. This term becomes smaller, and thus the bounds become tighter, when $\P$ assigns higher probability mass to controllers with lower costs. Designing such priors manually is challenging, so we aim to learn them from data. However, \autoref{corol:qstar} requires $\P$ to be independent of the dataset $\S$.

To address this, we split the dataset into two disjoint subsets, $\S = \S_1 \cup \S_2$ with $\S_1 \cap \S_2 = \varnothing$, and adopt a two-stage procedure. In the first stage, we use $\S_1$ together with the data-independent prior $\P_1 = \P$ to infer a posterior $\Q_1$. This posterior then serves as a data-dependent prior for the second stage, $\P_2 = \Q_1$, which uses $\S_2$ to infer the posterior $\Q_2$. The process is illustrated in \autoref{fig:twostage}. Posterior inference in both stages follows the procedure in \autoref{corol:qstar}.

To ensure that the two-stage and original single-stage procedures have the same performance, we require that they obtain the same posterior, i.e., $\Q_2 = \Q^*$. The following result establishes conditions under which this equivalence holds.

\begin{proposition}\label{prop:twostage}
    Let $\S_1$ and $\S_2$ be subsets of the dataset $\S$ with sizes $\numRollouts_1$ and $\numRollouts_2$ such that $\S_1 \cup \S_2 = \S$ and $\S_1 \cap \S_2 = \varnothing$. Define the posterior distributions in the two stages as Gibbs distributions with parameters $\tempGibbs_1 = \frac{\numRollouts_1}{\numRollouts} \, \tempGibbs$ and $\tempGibbs_2 = \frac{\numRollouts_2}{\numRollouts} \, \tempGibbs$:
    \begin{align*}
        \Q_1(\contParam) 
        =& \P_1(\contParam) \,
            e^{-{\tempGibbs_1} \costEmp(\contParam, \S_1)}
            / Z_{\tempGibbs_1}(\P_1, \S_1) \; , \quad \P_1=\P,
        \\
        \Q_2(\contParam) 
        =& \P_2(\contParam) \,
            e^{-{\tempGibbs_2} \costEmp(\contParam, \S_2)}
            / Z_{\tempGibbs_2}(\P_2, \S_2) \; , \quad \P_2=\Q_1.
    \end{align*}
    Then, $\Q_2$ coincides with the optimal posterior $\Q^*$ from \autoref{corol:qstar}, corresponding to the prior $\P$ and the full dataset $\S$.\looseness-1
\end{proposition}

The proof is given in \aref{app:proof_twostage}. Since $\Q_2 = \Q^*$, the two-stage procedure achieves the same performance as the single-stage approach. However, it offers an additional advantage: the bounds obtained by applying \autoref{corol:ub_qstar} in the second stage—with prior $\P_2$, parameter $\tempGibbs_2$, and dataset $\S_2$—can be tighter than those obtained in the single-stage case with prior $\P$, parameter $\tempGibbs$, and dataset $\S$. This improvement is confirmed empirically in \autoref{sec:experiments}.
 
\section{Approximate Inference}\label{sec:approx}
\begin{figure}
  \centering
  \includegraphics[width=\linewidth]{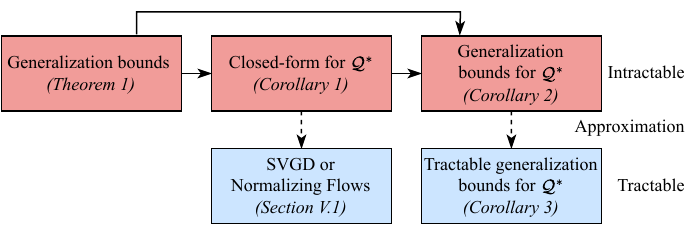}
  \caption{Overview of the components in our framework, highlighting their role in building a tractable and theoretically grounded pipeline. The approximation methods introduced in \autoref{sec:approx} replace intractable components (\colorbox{intractableCol}{\phantom{e}}) with tractable counterparts (\colorbox{tractableCol}{\phantom{e}}).
  }
  \label{fig:relations}
\end{figure}

We introduced the optimal posterior in \autoref{corol:qstar} and its associated generalization bounds in \autoref{corol:ub_qstar}. However, their direct use is limited by the partition function $Z_\tempGibbs$ in~\eqref{eq:Z}, which is generally intractable due to the expectation over $\contParam \sim \P$. In this section, we develop approximation methods that make these results practically applicable. 
We first approximate $\Q$ with a tractable distribution from which controllers $\contParam$ can be efficiently sampled. We then approximate the function $Z_\tempGibbs$ to obtain computable bounds on the true cost.
An overview of the relationships between the theoretical components of our framework is illustrated in \autoref{fig:relations}.

\subsection{Approximating the optimal posterior}\label{subsec:approx_Q}
Given $\P$ and $\S$, the optimal posterior $\Q^*$ is only computable via~\eqref{eq:qstar} up to the normalization constant $Z_\tempGibbs$. Hence, direct sampling from $\Q^*$ is infeasible. To address this, we approximate $\Q^*$ with a tractable distribution from which we can sample the controller parameters $\contParam$.
We employ two approximation techniques: SVGD~\cite{liu2016stein} and normalizing flows~\cite{rezende2015variational}. 

\begin{figure}
  \centering
  \includegraphics[width=\linewidth]{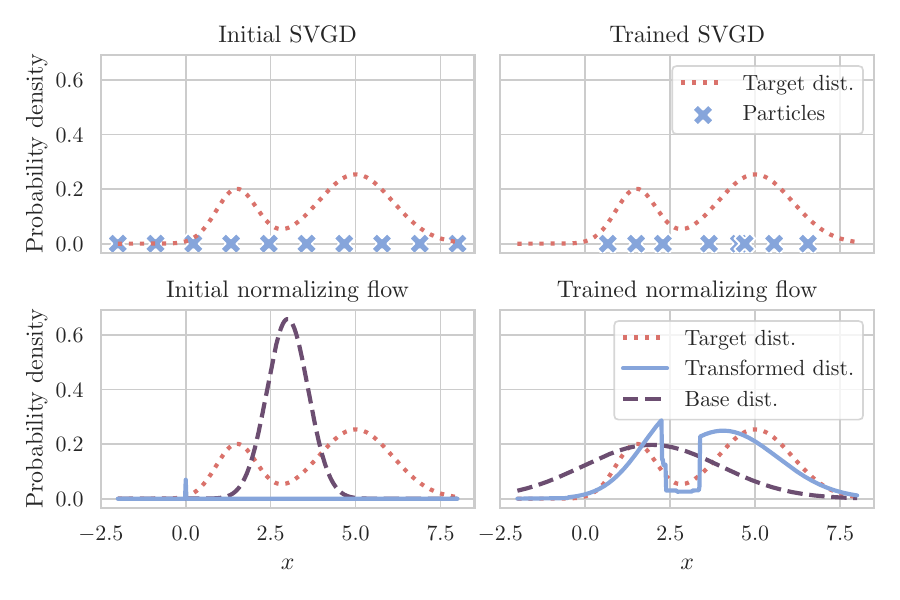}
  \caption{Approximating a one-dimensional target distribution (red) over a variable $x$ using SVGD (top) and normalizing flows (bottom). In SVGD, particles (cyan crosses) are initialized uniformly and move toward high-density regions after training. In normalizing flows, both the Gaussian base distribution (magenta) and the transformations are trainable, allowing the transformed distribution (cyan) to better align with the target after training.}
  \label{fig:svgd_1d}
\end{figure}

SVGD approximates $\Q^*$ using a set of $\numParticles \in \mathbb{N}$ particles denoted by ${\particle_1, \dots, \particle_\numParticles}$. In our framework, each particle corresponds to the parameters of a controller, i.e., $\phi_\counterParticle \in \contParamSet$ for $\counterParticle \in \{ 1, \dots, \numParticles\}$. SVGD initializes these particles by sampling from the prior distribution and iteratively updates them to align their density with the probability mass of $\Q^*$ through gradient descent.
After training, the approximated distribution is a uniform distribution over the particles, making sampling equivalent to randomly picking a particle with equal probability. A one-dimensional illustration of SVGD particles before and after training is provided in \autoref{fig:svgd_1d}.
We refer to~\cite{liu2016stein} for further details.


Although SVGD can approximate complex distributions with arbitrary precision given sufficient particles~\cite{liu2016stein}, its computational cost grows linearly with the number of particles. To balance accuracy with efficiency, we use normalizing flows~\cite{rezende2015variational}. This approach transforms an initial simple density, called the \textit{base} distribution, into a complex one through parameterized invertible mappings. The parameters of these invertible mappings and the base distribution are updated through gradient descent to align the resulting transformed distribution with $\Q^*$. A one-dimensional example of normalizing flows before and after training is provided in \autoref{fig:svgd_1d}. 
We empirically compare SVGD and normalizing flows in \autoref{sec:experiments}.

\subsection{Approximating the bound}\label{subsec:approx_ub}
Next, we tackle the intractability of the upper bound for the optimal posterior in \autoref{corol:ub_qstar}, which arises due to its dependence on the partition function, $Z_\tempGibbs$. To address this, we approximate $\ln Z_\tempGibbs$ via Monte Carlo methods and account for potential overestimation using McDiarmid's inequality~\cite{McDiarmid_1989}. This ensures a valid, albeit looser, upper bound, as established in the following corollary.

\begin{corollary}\label{corol:mcdim_ub}
    Let $\contParam_1, \dots, \contParam_{\numPSamples}$ be  $\numPSamples\in\N$ independent samples from the prior $\P$, and define the Monte Carlo estimate:
    \begin{align}
        \hat{Z}_\tempGibbs(\contParam_1, \cdots, \contParam_{\numPSamples}, \S) 
        =
        \frac{1}{\numPSamples} \sum_{\counterPSamples=1}^{\numPSamples}
            e^{-\tempGibbs  \,\costEmp(\contParam_{\counterPSamples}, \S)}
        . 
        \label{eq:Zhat}
    \end{align}
    If the optimal posterior $\Q^*$ from~\eqref{eq:qstar}-\eqref{eq:Z} is used in \autoref{theo:pac}, as per \autoref{corol:ub_qstar}, the inequalities,
    \begin{align}
        \costTrue &(\contParam, \D_{T:0}) 
        \leq
        - \frac{1}{\tempGibbs} \ln \hat{Z}_\tempGibbs(\contParam_1, \cdots, \contParam_{\numPSamples}, \S) 
        + \frac{1}{\tempGibbs} \ln \frac{1}{\confidence}
        \nonumber \\
        & 
        + \frac{\tempGibbs \costBound^2}{8 \numRollouts} 
        + \frac{1}{\tempGibbs} \ln \Bigl(1+ \frac{e^{\tempGibbs \costBound} - 1}{\numPSamples} \Bigr) \sqrt{\frac{\numPSamples}{2} \ln\bigl(\frac{1}{\confidence}\bigr)},
        \label{eq:u_bound_qstar_mcdim}
        \\
        \costTrue&(\contParam, \D_{T:0})
        \geq   
        2 \costEmp(\contParam, \S)
        +
        \frac{1}{\tempGibbs} \ln \hat{Z}_\tempGibbs(\contParam_1, \cdots, \contParam_{\numPSamples}, \S) 
        - 
        \frac{1}{\tempGibbs} \ln \frac{1}{\confidence}
        \nonumber \\
        &
        - 
        \frac{\tempGibbs \costBound^2}{8 \numRollouts}
        - \frac{1}{\tempGibbs} \ln \Bigl(1+ \frac{e^{\tempGibbs \costBound}-1}{\numPSamples} \Bigr) \sqrt{\frac{\numPSamples}{2} \ln\bigl(\frac{1}{\confidence}\bigr)}, 
        \label{eq:l_bound_qstar_mcdim}
    \end{align}
    hold individually with probability at least $1-2\,\confidence$ and jointly with probability at least $1-3\,\confidence$. The probability is over simultaneously sampling $\S \sim \D_{T:0}^\numRollouts$, $\contParam \sim \Q^*$, and $\contParam_1, \cdots, \contParam_{\numPSamples} \sim \P^{\numPSamples}$.
\end{corollary}
The proof is provided in \autoref{app:proof_mcdim_ub}.

The last term on the right-hand side of \eqref{eq:u_bound_qstar_mcdim} and \eqref{eq:l_bound_qstar_mcdim} reflects the additional conservatism introduced by the approximation. This term decreases as the number of samples $\numPSamples$ increases, ultimately vanishing as $\numPSamples\to \infty$. Moreover, since $\ln \hat{Z}_\tempGibbs$ is an unbiased estimator of $\ln Z_\tempGibbs$, \autoref{corol:ub_qstar} and \autoref{corol:mcdim_ub} are consistent and coincide in the limit $\numPSamples \to \infty$.

A related approximation is provided in~\cite{Boroujeni2024PACSNOC} by bounding the difference between $Z_\tempGibbs$ and $\hat{Z}_\tempGibbs$ for sufficiently large $\numPSamples$. In contrast, we bound the deviation between $\ln Z_\tempGibbs$ and $\ln \hat{Z}_\tempGibbs$. Compared to~\cite{Boroujeni2024PACSNOC}, our relaxed bound decays faster with $\numPSamples$ and holds for any $\numPSamples$, as demonstrated in our experiments in \autoref{subsec:experiments_robots}.
\section{Experiments}\label{sec:experiments}
We evaluate the proposed method on two systems:
(i) a scalar Linear Time-Invariant (LTI) system, and
(ii) a multi-agent navigation task in which two planar robots must reach predefined target locations while avoiding collisions.
For both systems, the posterior distribution over controller parameters is given by the optimal Gibbs distribution $\mathbf{Q}^*$ with parameter $\tempGibbs^*$, as defined in \eqref{eq:qstar} and \eqref{eq:lambdastar}, respectively. In all experiments, we apply the cost transformation in \eqref{eq:cost_trans} with $\costBound = 1$, and set $\costDen$ as discussed in \autoref{subsec:PAC4SNOC}. 

The source code for all experiments is available at \url{https://www.github.com/DecodEPFL/PAC-SNOC}. 
The implementation is based on PyTorch~\cite{pytorch} modules and supports batched computations for efficient GPU usage. 
All experiments were conducted on an NVIDIA H100 GPU, with each run restricted to $40\%$ of the available GPU resources.

\subsection{LTI system}
\paragraph*{\textbf{System}}
We consider a scalar, stable LTI system to illustrate the control design procedure and the derived PAC-Bayesian bounds. The system is given by:
\begin{align*}
    \state_0 &= \stateInitNom + \noise_0,\\ 
    \state_t &= a \, \state_{t-1} + b \, \inp_{t-1} + \noise_t , \quad t=1,2, \ldots,
\end{align*}
with $a = 0.8$, $b = 0.1$, and $\stateInitNom = 2$. At each time step, the noise $\noise_t$ is drawn from the Gaussian distribution $\mathcal{N}(\mu_\noise, \sigma_\noise^2)$ with $\mu_\noise = 0.3$ and $\sigma_\noise^2 = 0.09$. The distribution of $\noise_t$ is assumed unknown; instead, we are given $\numRollouts$ independent noise sequences of length $\horizon = 10$. The value of $\numRollouts$ is varied to study its influence on the performance.

\paragraph*{\textbf{Controller}} We utilize the affine state-feedback controller $u_t = -(k \, x_t + \beta)$, where $k$ and $\beta$ denote the controller gain and bias. The vector $\theta = [k, \beta]^\top$ represents the controller parameters. Closed-loop stability corresponds to the constraint $k \in (-2, 18)$, which we enforce during optimization by projecting $k$ into this interval at each update. Because the system is scalar and LTI, this simple projection step alone ensures stability, obviating the need for more elaborate stability certification methods such as those in \autoref{subsec:controller_parametrization}.
The controller parameters are optimized to minimize the quadratic FH cost $\costTraj(\state_{\horizon:0}, \inp_{\horizon:0}) = \sum_{\timestep=0}^\horizon 5 \, \state_\timestep^2 + 0.003 \, \inp_\timestep^2$.

\paragraph*{\textbf{Baselines}}
We compare our methodology against two alternatives. The first is an \textit{empirical} controller obtained by minimizing $\costEmp(\theta, \S)$ as defined in~\eqref{eq:emp_cost}. The second is a \textit{benchmark} (ideal) controller that assumes exact knowledge of the noise mean $\mu_\noise$, enabling the optimal choice $\beta = \mu_\noise / b$. The gain $k$ of the benchmark controller is optimized by minimizing the empirical cost over a large dataset of $1024$ noise sequences. While this controller achieves strong performance, it relies on both accurate knowledge of $\mu_\noise$ and a substantial amount of data, which limits its practicality. We nonetheless include it as a best-case competitor.

\paragraph*{\textbf{Prior distribution}}
The prior distribution over the controller parameters $\theta = [k, \beta]^\top$ must be selected independently of the dataset $\S$. We choose independent priors for $k$ and $\beta$. Specifically, we set the prior for $k$ as a Gaussian distribution centered at the infinite-horizon LQR (IH-LQR) gain with variance $1.0$. Although the IH-LQR gain is not necessarily optimal for the finite horizon $\horizon = 10$, it provides a reasonable, data-independent choice.
We select the prior for $\beta$ to reflect available knowledge of the noise mean $\mu_\noise$. We consider two cases:
(i) when it is only known that $\mu_\noise \in [-0.5, 0.5]$, we set $\beta \sim \mathcal{U}(-0.5/b, 0.5/b) = \mathcal{U}(-5, 5)$;
(ii) when $\mu_\noise$ is known with some uncertainty, we set $\beta \sim \mathcal{N}(\mu_\noise / b, 1.5^2) = \mathcal{N}(3, 1.5^2)$.
We refer to these priors as $\P_\mathcal{U}$ and $\P_\mathcal{N}$, respectively.

\paragraph*{\textbf{Posterior distribution}}
Since both $\P$ and $\Q^*$ are bivariate in the controller parameters $(k, \beta)$, we approximate their Probability Density Functions (PDFs) by discretizing a rectangular grid in the $(k, \beta)$-plane that contains more than $95\%$ of the probability mass. In two dimensions, such gridding is not computationally expensive; therefore, we adopt this approach instead of more complex sampling-based approximations such as SVGD or normalizing flows. Figure~\ref{fig:Gibbs_grid} shows $\P$ after discretization in the left column, and the discretized $\Q^*$ together with the empirical and benchmark controllers in the middle and right columns for $\confidence = 0.2$. The middle column corresponds to $S = 8$, whereas the right column corresponds to $S = 512$.
The top and bottom rows correspond to $\P_\mathcal{U}$ and $\P_\mathcal{N}$, respectively, and the color encodes the approximated PDF. The empirical and benchmark controllers are also indicated.

\paragraph*{\textbf{Evaluating the performance}}
For $\numRollouts = 8$, the posterior distributions associated with both $\P_\mathcal{U}$ and $\P_\mathcal{N}$ are concentrated near the benchmark controller, whereas the empirical controller lies farther away. This suggests that controllers sampled from the posterior are more likely to resemble the benchmark controller, potentially yielding superior out-of-sample performance compared to the empirical controller. With $\numRollouts$ increased to $512$, each prior yields a posterior more concentrated around the benchmark controller. Likewise, the empirical controller also improves and approaches the benchmark.

A comparison of the two rows of Figure~\ref{fig:Gibbs_grid} shows that the more informative prior $\P_\mathcal{N}$ (bottom row), which incorporates additional knowledge of $\mu_\noise$, produces a sharper posterior distribution. This confirms that the proposed algorithm effectively integrates prior knowledge into the control design process.

\begin{figure}
  \centering
  \vspace{1pt}
  \includegraphics[width=0.96\linewidth]{figs/N_biased_Gaussian_biased_wide_contour_T10_S8_512.pdf}
  \caption{Discretized PDFs of the prior distributions (left) and the optimal posterior distributions for $\numRollouts = 8$ (middle) and $\numRollouts = 512$ (right). The top and bottom rows correspond to $\P_\mathcal{U}$ and $\P_\mathcal{N}$, respectively. In each plot, the horizontal and vertical axes represent $\beta$ and $k$, respectively, and the color encodes the PDF value. The empirical and benchmark controllers are indicated by markers.}
  \label{fig:Gibbs_grid}
\end{figure}

\paragraph*{\textbf{Evaluating the bounds}}
A desirable upper bound should ensure that the true cost of sampled controllers remains below the bound in at least $1 - \confidence$ of the cases, while avoiding excessive looseness. We evaluate the approximate bound in~\autoref{corol:mcdim_ub} when using $\tempGibbs^*$ under varying training set size $\numRollouts$, prior distribution $\P$, and confidence level $\confidence$. For each configuration, we compute $\Q^*$ and $\tempGibbs^*$, draw $10$ controller parameters from the resulting posterior distribution, and compute the true cost of each controller using exhaustive test data sampling.

\begin{figure}
  \centering
  \includegraphics[width=0.99\linewidth]{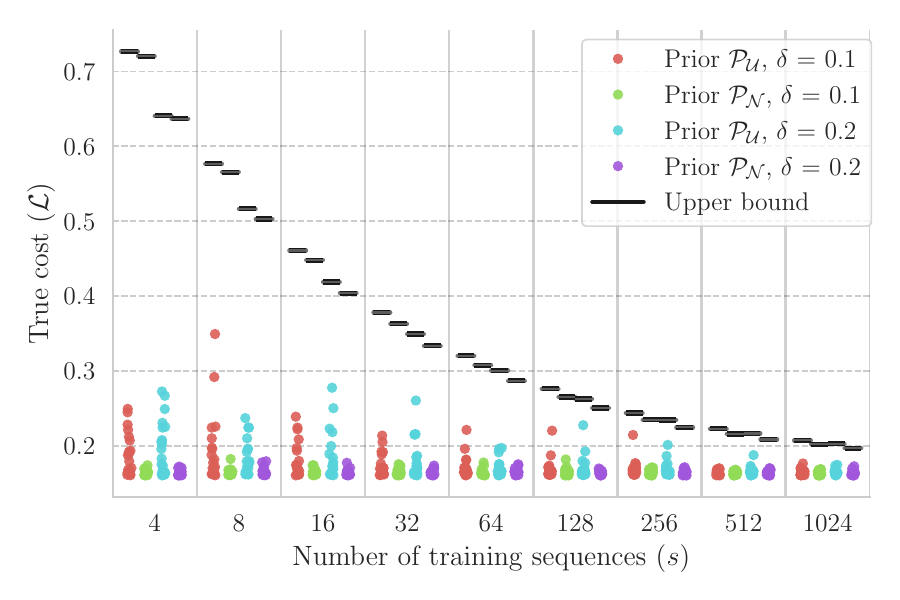}
  \caption{Comparison of the transformed true cost $\costTrue$ and the upper bound in \autoref{corol:ub_qstar} for various configurations as a function of $\numRollouts$. Colors indicate the choices of $\confidence$ and prior distribution $\P$. For each configuration, the true cost is approximated for $10$ parameter vectors $\theta$ sampled from $\Q^*$, shown as vertically aligned circles.}
  \label{fig:ub}
\end{figure}

\autoref{fig:ub} illustrates the cost of each sampled controller on the test data (shown as circles) together with the corresponding upper bounds from \autoref{corol:ub_qstar}, computed by discretizing the prior distribution (black line). The horizontal axis shows different values of $\numRollouts$, and colors indicate the choice of confidence level $\confidence$ and prior distribution $\P$. Results are shown for REN-based controllers, while experiments with SSMs lead to analogous conclusions.

In Figure~\ref{fig:ub}, the transformed true costs consistently lie below the corresponding upper bounds, confirming their validity. For fixed $\confidence$ and $\numRollouts$, the bound for $\P_\mathcal{N}$ is tighter than that for $\P_\mathcal{U}$, consistent with the discussion in Section~\ref{subsec:qstar} that a more informative prior yields a smaller bound. For a fixed prior and $\numRollouts$, increasing $\confidence$ tightens the bound by allowing a higher violation probability. Finally, increasing $\numRollouts$ uniformly tightens the bound across all settings. A formal theoretical analysis of the dependence on $\numRollouts$ is left for future work.

\subsection{Navigating planar robots}\label{subsec:experiments_robots}

\begin{table*}[t]
\caption{Performance and training time for the empirical approach and PAC-SNOC with $\numRollouts = 32$. PAC-SNOC results are reported for REN and SSM architectures, with the posterior approximated using SVGD or normalizing flows. Metrics include the transformed cost (via~\eqref{eq:cost_trans} with $\costBound = 1$) on training and test sets, the collision rate (percentage of trajectories with robot–robot or robot–obstacle collisions, counting multiple collisions within a trajectory), and training time until reaching convergence or $5000$ epochs. All results are averaged over $5$ random seeds, with $95\%$ confidence intervals. PAC-SNOC consistently outperforms the empirical approach on the test set in both cost and collision rate. Best values for each criterion and architecture are highlighted in green.}
\label{tab:robots}
\begin{center}
\begin{tabular}{cccccccc}
\toprule
\multicolumn{1}{c}{} & \multicolumn{1}{c}{} & \multicolumn{1}{c}{} & \multicolumn{2}{c}{\textbf{Cost}} & \multicolumn{2}{c}{\textbf{Collisions Percentage}} & \multicolumn{1}{c}{} \\
\cmidrule(rl){4-5} \cmidrule(rl){6-7}
\textbf{Method} & \textbf{NN type} & \textbf{Inference} & {Train} & {Test} & {Train} & {Test} & \textbf{Training Time (s)}\\
\midrule
Empirical & REN & - & \cellcolor{\bestres}$0.0866 \pm 0.0059$ & $0.0921 \pm 0.0096$ & \cellcolor{\bestres}$0.25 \pm 0.49$ & $5.27 \pm 6.16$ & \cellcolor{\bestres}$3083 \pm 942$\\
PAC-SNOC & REN & SVGD - 1 particle & $0.0871 \pm 0.0011$ & $0.0881 \pm 0.0014$ & $0.63 \pm 0.25$ & $0.90 \pm 0.02$ & $3348 \pm 987$\\
PAC-SNOC & REN & SVGD - 5 particles & $0.0873 \pm 0.0014$ & \cellcolor{\bestres}$0.0731 \pm 0.0008$ & $0.78 \pm 0.31$ & \cellcolor{\bestres}$0.69 \pm 0.03$ & $16680 \pm 3200$\\
PAC-SNOC & REN & normalizing flows & $0.0869 \pm 0.0019$ & $0.0744 \pm 0.0015$ & $0.00 \pm 0.00$ & $0.72 \pm 0.09$ & $5495 \pm 2504$\\ 
\midrule
Empirical & SSM & - & \cellcolor{\bestres}$0.0857 \pm 0.0016$ & $0.0904 \pm 0.0031$ & \cellcolor{\bestres}$0.00 \pm 0.00$ & $4.64 \pm 0.16$ & \cellcolor{\bestres}$4200 \pm 1184$\\
PAC-SNOC & SSM & SVGD - 1 particle & $0.0875 \pm 0.0009$ & $0.0882 \pm 0.0016$ & $0.67 \pm 0.23$ & $0.91 \pm 0.02$ & $4600 \pm 962$\\
PAC-SNOC & SSM & SVGD - 5 particle & $0.0890 \pm 0.0008$ & \cellcolor{\bestres}$0.0727 \pm 0.0010$ & $0.75 \pm 0.29$ & \cellcolor{\bestres}$0.65 \pm 0.01$ & $22516 \pm 2874$\\
PAC-SNOC & SSM & normalizing flows & $0.0881 \pm 0.0017$ & $0.0761 \pm 0.0021$ & $0.03 \pm 0.02$ & $0.75 \pm 0.03$ & $6214 \pm 1320$\\
\bottomrule
\end{tabular}
\end{center}
\end{table*}

\begin{figure}
    \centering
    \begin{minipage}[t]{0.33\linewidth}
        \includegraphics[width=\linewidth]{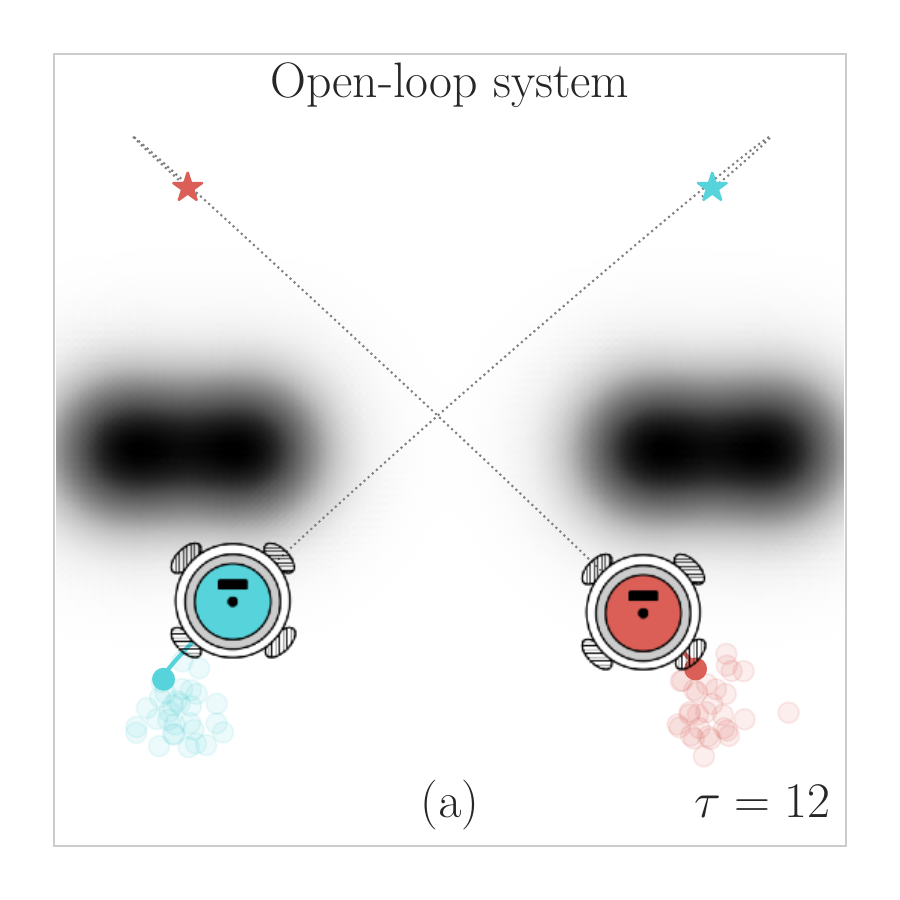}
        \includegraphics[width=\linewidth]{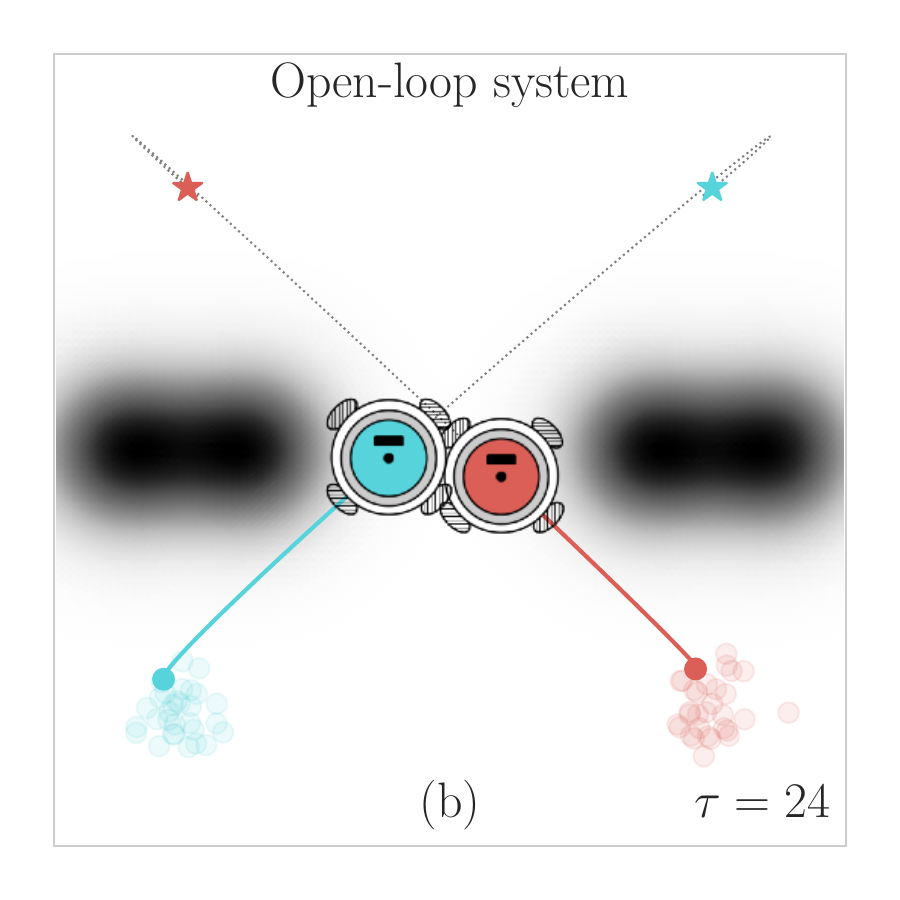}
    \end{minipage}%
    \begin{minipage}[t]{0.33\linewidth}
        \includegraphics[width=\linewidth]{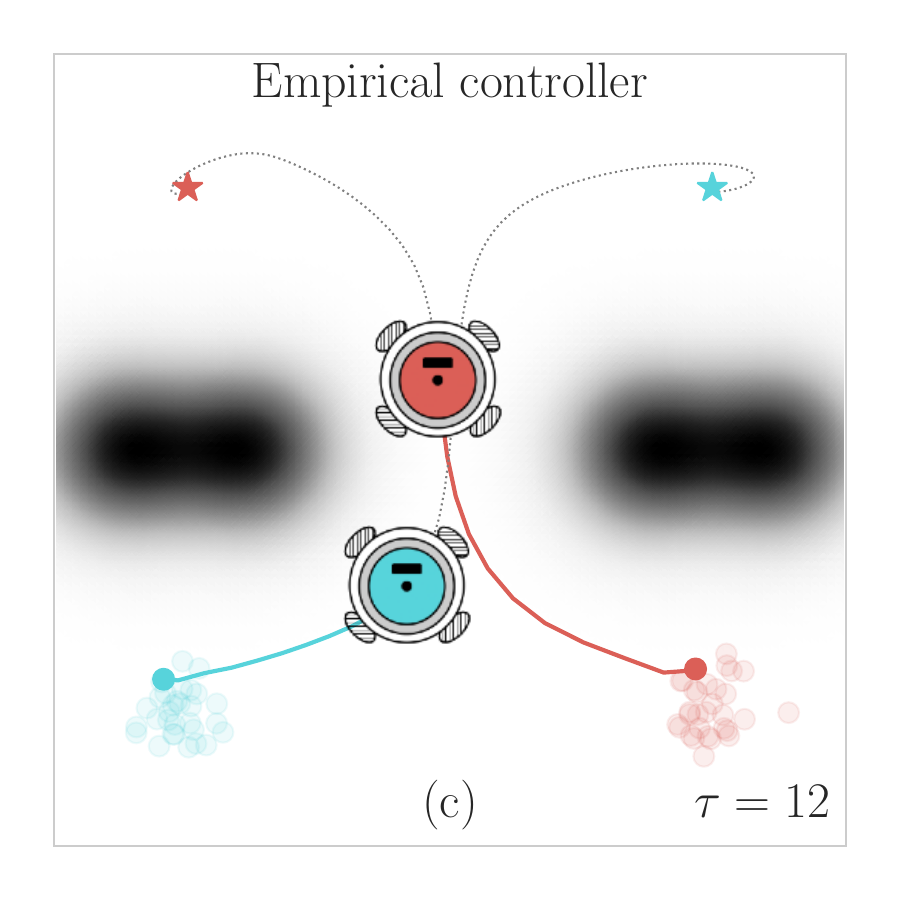}
        \includegraphics[width=\linewidth]{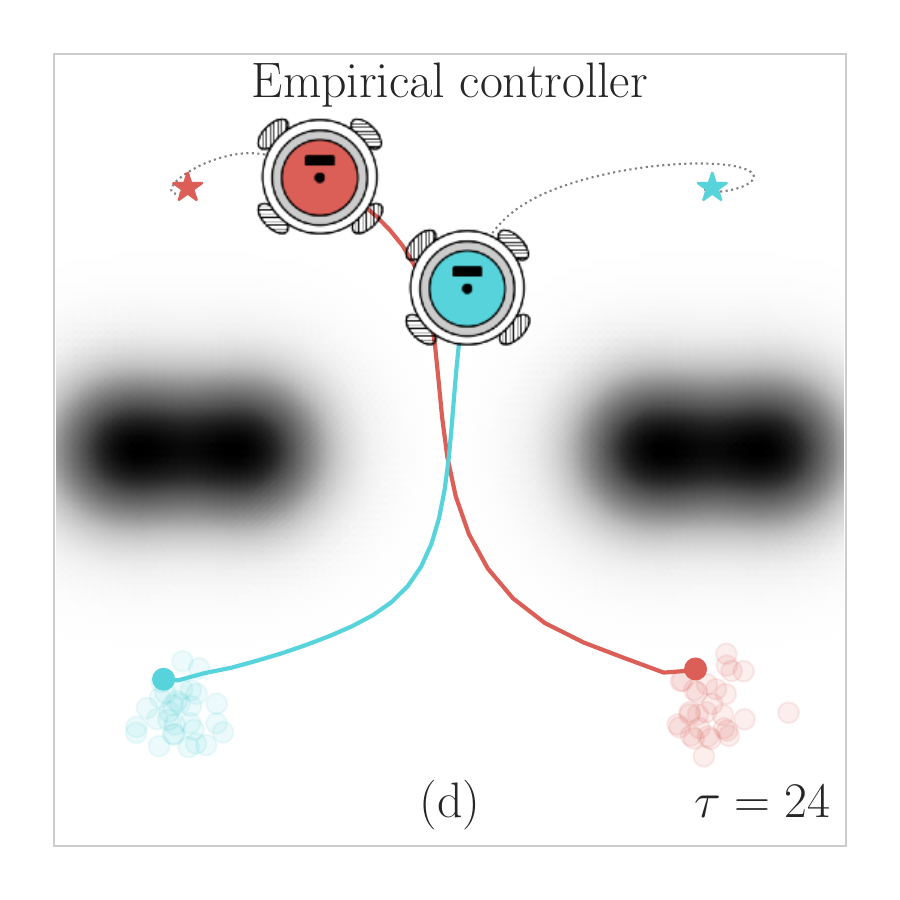}
    \end{minipage}%
    \begin{minipage}[t]{0.33\linewidth}
        \includegraphics[width=\linewidth]{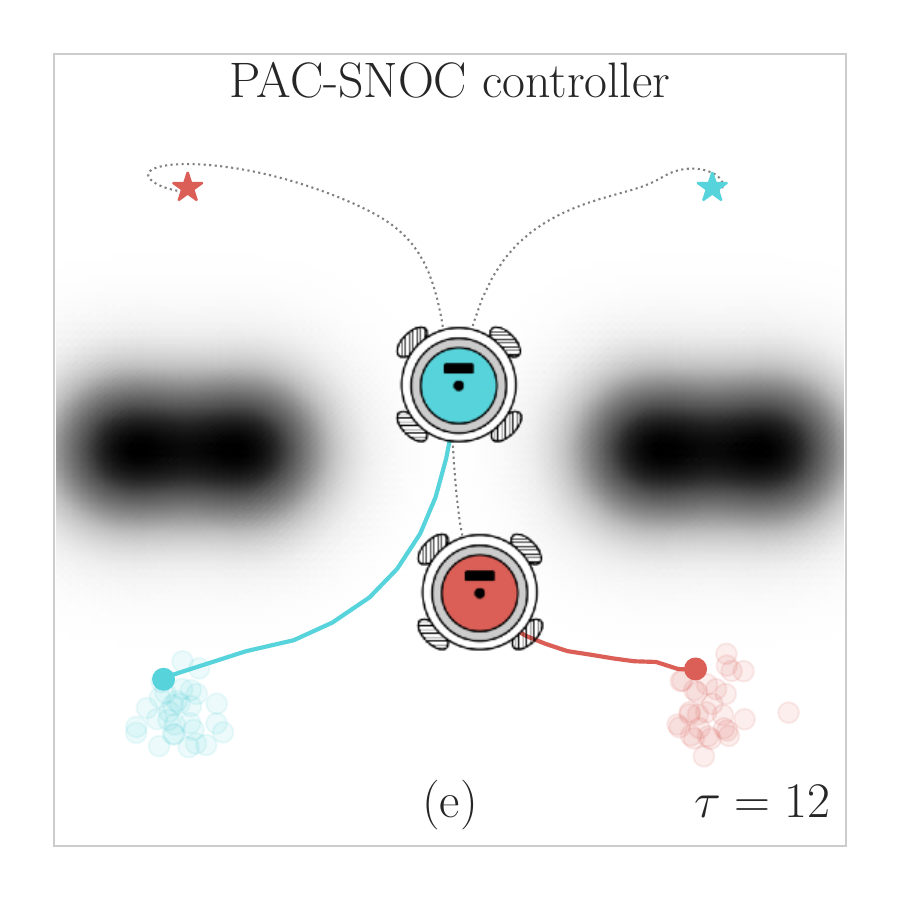}
        \includegraphics[width=\linewidth]{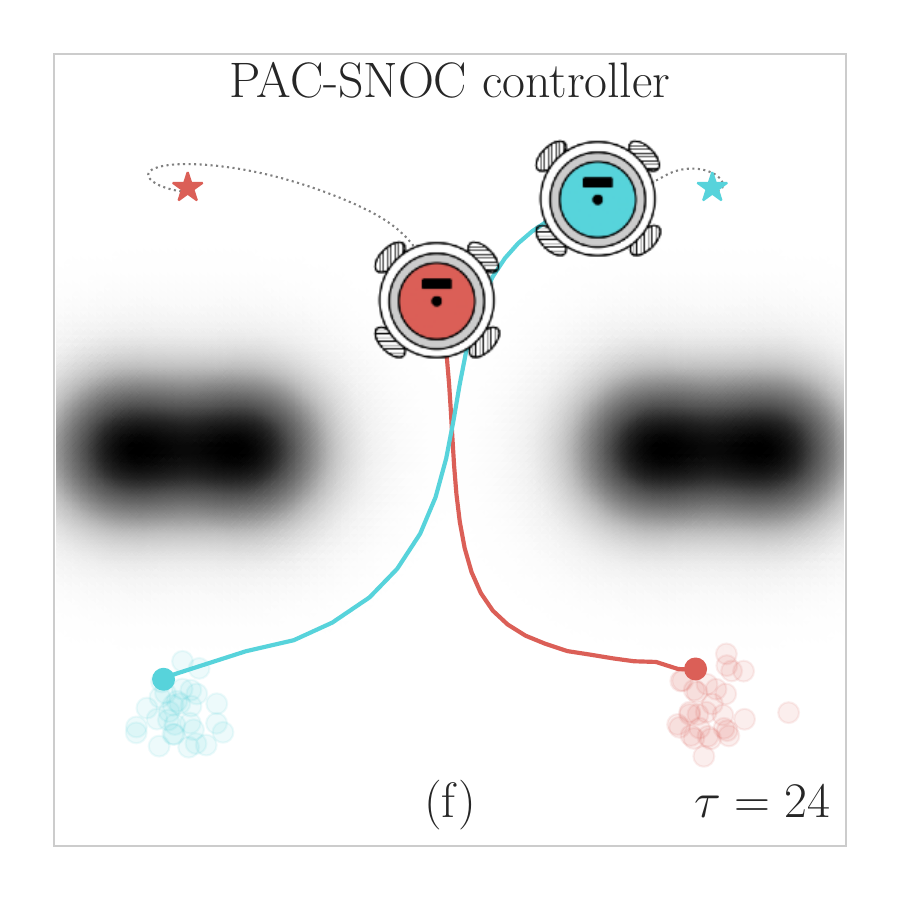}
    \end{minipage}
    \caption{
    Closed-loop test trajectories of (\textit{a})–(\textit{b}) the pre-stabilized system, (\textit{c})–(\textit{d}) the empirical controller, and (\textit{e})–(\textit{f}) a controller sampled from $\Q^*$ using SVGD.
    Training and test initial conditions are indicated by transparent and solid circles, respectively. Snapshots are taken at the instants $\tau$ shown in each plot. Solid lines depict the trajectory from $t=0$ to $\tau$, while dotted lines show the continuation until $t=400$.
    } 
\label{fig:corridor}
\end{figure}

\paragraph*{\textbf{System}}
We consider two robots coordinately passing through a narrow valley and reaching a target position while avoiding both obstacles and mutual collisions—see Figure~\ref{fig:corridor}. Each robot is modeled as a point-mass double integrator subject to nonlinear drag forces. The overall system state is $\state \in \R^8$, comprising the positions and velocities of both robots in the Cartesian coordinates, and the control input $\inp \in \R^4$ represents the forces applied to each robot along the two coordinate directions. Each robot is prestabilized by a proportional controller that drives it toward the desired end position but lacks explicit collision-avoidance logic, which can result in collisions and degraded performance. This experiment is adapted from~\cite{NeurSLS}, where additional details on the system setup are provided.

The noise $\noise_\timestep$ in~\eqref{eq:system} affects only the initial condition, i.e., $\noise_\timestep = 0$ for $\timestep > 0$, with $\noise_0$ drawn from a zero-mean Gaussian distribution of variance $\sigma_\noise^2 = 0.2^2$. The value of $\sigma_\noise$ is assumed to be unknown. The dataset consists of $\numRollouts = 32$ independent noise sequences over a horizon of $\horizon = 100$, with nonzero values only at $\timestep = 0$.

\paragraph*{\textbf{Controller}}
Given the complexity of the task, we employ NN-based controllers following the stability-preserving architecture described in~\autoref{subsec:controller_parametrization}. The mapping $\Emme_{\infty:0}^{\contParam}$, which is required to be $\lp$-stable, is implemented in two NN variants: one based on RENs and the other on SSMs. The dimensions of the REN and SSM are chosen so that each contains approximately $900$ parameters, enabling a fair performance comparison. Further details on these models are provided in~\aref{app:implementation}.

The controller parameters are optimized to minimize an FH cost comprising a quadratic term plus collision- and obstacle-avoidance penalties:
\begin{equation*}
    \costTraj(\state_\timestep, \inp_\timestep) = \sum_{\timestep=0}^\horizon (\delta \state_\timestep)^\top Q (\delta \state_\timestep) + \inp_\timestep^\top R \inp_\timestep
+ l_{ca}(d_t) + l_{oa}(\state_\timestep),
\end{equation*}
where $\delta \state_\timestep = \state_\timestep - \state_{\text{target}}$, $\state_{\text{target}}$ is the desired state with target positions and zero velocities, and $d_\timestep \in \R_0^+$ is the distance between robots at time $\timestep$. The collision-avoidance term $l_{ca}(d_\timestep)$ is zero if $d_\timestep > D$ for a safe distance $D > 0$, and otherwise $l_{ca}(d_\timestep) = (d_\timestep + \nu)^{-2}$ with $\nu > 0$ a small positive constant. The obstacle-avoidance term $l_{oa}$ is a function of the distance between the robots and each obstacle.
The exact formulation of the cost function is provided in~\cite{NeurSLS}.

\paragraph*{\textbf{Baselines}}
We compare our method against the empirical approach, which minimizes $\costEmp(\theta, \S)$ for each controller architecture. No benchmark approach is considered in this experiment, as the large number of controller parameters would require an impractically large dataset to train an ideal high-performing controller, and manually incorporating knowledge of the noise distribution into the controller design procedure is challenging.

\paragraph*{\textbf{Prior distribution}}
We use a zero-mean spherical Gaussian prior on the parameters $\contParam$ of the controller---whether implemented as a REN or as an SSM--- which imposes L$_2$-regularization over $\contParam$~\cite{Bishop2006}. This choice is motivated by the well-established effectiveness of L$_2$-regularization in mitigating overfitting in supervised learning. The variance of the prior is fine-tuned via Bayesian optimization (see~\aref{app:optuna}).

\paragraph*{\textbf{Posterior distribution}}
We approximate the optimal Gibbs posterior distribution $\Q^*$, corresponding to $\tempGibbs = \tempGibbs^*$ with confidence level $\confidence = 0.1$, using the SVGD and normalizing flows methods described in~\autoref{subsec:approx_Q}. For SVGD, we evaluate configurations with $1$ and $5$ particles to examine the trade-off between computational cost and approximation quality. For normalizing flows, we employ a Gaussian base distribution followed by a sequence of $16$ planar transformations, as introduced in~\cite{rezende2015variational}. Further implementation details are provided in~\aref{app:implementation}.

\paragraph*{\textbf{Evaluating the performance}}
\autoref{tab:robots} compares the performance and training time of the empirical approach with the proposed PAC-SNOC method using a dataset of $\numRollouts = 32$ noise sequences, split into $75\%$ for training and $25\%$ for validation. We evaluate both REN and SSM controller architectures and, for PAC-SNOC, approximate the posterior distribution using either SVGD or normalizing flows. In \autoref{tab:robots}, we report the cost on both the training set and an unseen test dataset of $500$ noise trajectories. We also present the percentage of trajectories in which the robots collide with each other or with an obstacle, counting multiple collisions within the same trajectory. The training time is measured until reaching convergence or a maximum of $5000$ epochs. Convergence is defined as the absence of a substantial reduction in the validation cost for $500$ consecutive epochs. All experiments are repeated with $5$ different random seeds, and $95\%$ confidence intervals are reported. Best values for each criterion and architecture are marked in green.

The empirical controller achieves the lowest cost on the training set, while PAC-SNOC consistently outperforms it on the test set for all inference methods, both in terms of cost and collision rate. This confirms that the empirical controller overfits the training data, whereas PAC-SNOC regularizes the controllers. As a result, training performance is slightly worse, but test performance is substantially improved.

\begin{figure}
    \centering
    \includegraphics[width=\linewidth]{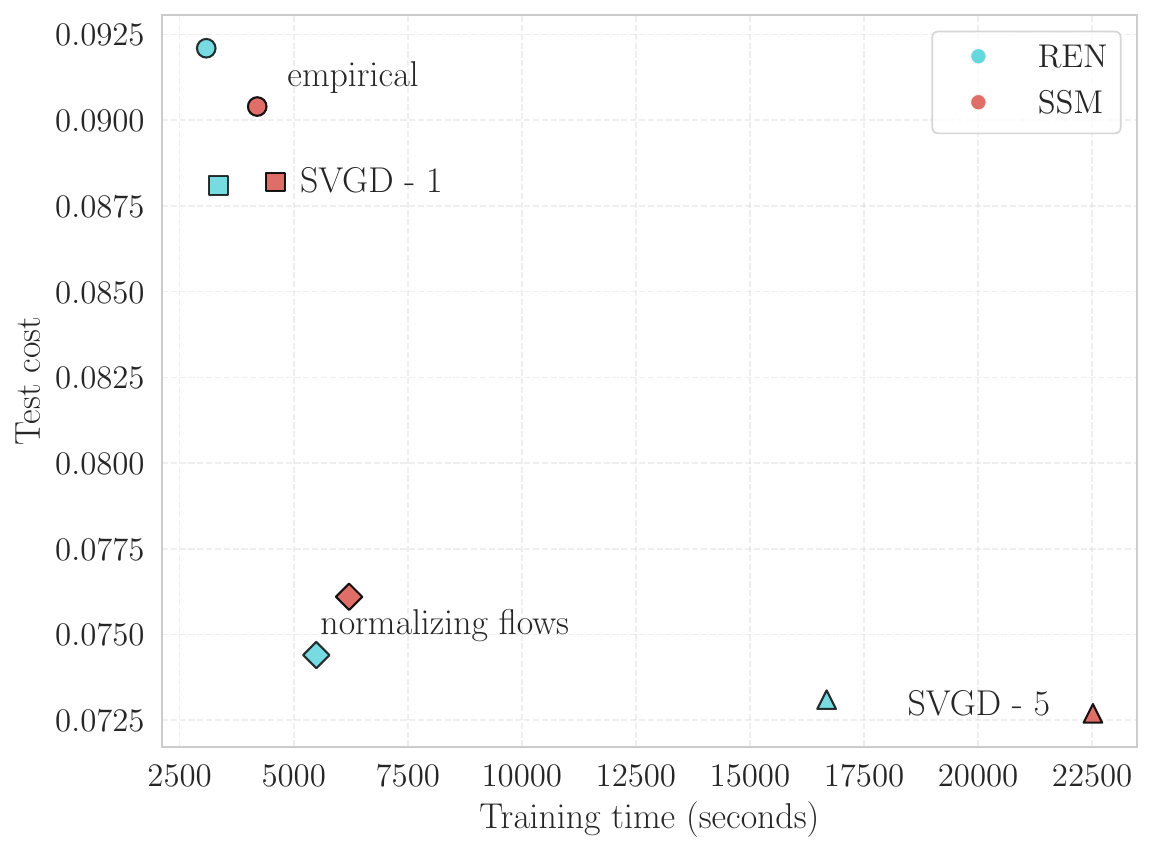}
    \caption{Trade-off between average test performance and training time with $\numRollouts = 32$ across random seeds. 
    SVGD with $5$ particles achieves the best performance but requires substantially more computation time. Normalizing flows perform slightly worse, but the training is significantly faster, offering a favorable accuracy–efficiency balance.}
\label{fig:cost_time}
\end{figure}

Among the inference methods, SVGD with $5$ particles achieves the best performance, but it is computationally intensive. Normalizing flows perform only slightly worse while requiring substantially less computation time, making them the preferred choice in practice. Across architectures, SSM slightly outperforms REN in most cases, albeit with a marginally higher computational cost. To further analyze the trade-off between performance and training time, we present \autoref{fig:cost_time}, which shows that SVGD with $5$ particles dominates in test performance but is slower to train, whereas normalizing flows offer a favorable balance between accuracy and efficiency.

Next, we visually compare the trajectories induced by the empirical and PAC-SNOC controllers in \autoref{fig:corridor}, using $\numRollouts = 32$, the REN architecture, and a fixed random seed. For PAC-SNOC, the posterior distribution is approximated using SVGD with $1$ particle. Each plot shows three trajectories starting from distinct initial conditions outside the training dataset, simulated beyond the horizon $\horizon$ to confirm that stability is preserved. The subplots correspond to: (\textit{a}) the prestabilized system without an external controller, (\textit{b})–(\textit{c}) the empirical controller, and (\textit{d})–(\textit{e}) the PAC-SNOC controller.
In subplot (\textit{a}), the prestabilizing controller causes collisions between robots and oscillations around the target position, indicating poor performance. In contrast, both the empirical and PAC-SNOC controllers in (\textit{b})–(\textit{e}) avoid collisions and yield smoother trajectories. Animations of these experiments are available in our \href{https://www.github.com/DecodEPFL/PAC-SNOC}{Github repository}.

\begin{figure}
  \centering
  \includegraphics[width=0.99\linewidth]{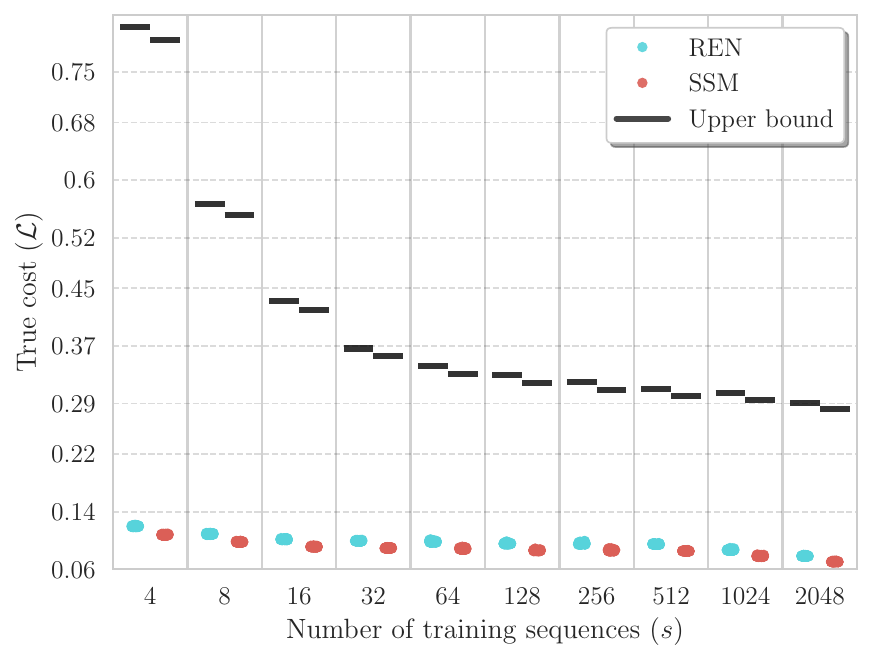}
  \caption{Test cost of $100$ controllers sampled from the posterior (circles) together with the corresponding upper bounds (black lines) computed using the tractable bound from \autoref{corol:mcdim_ub}. The horizontal axis shows the number of training noise sequences $\numRollouts$, while colors distinguish between REN- and SSM-based controller architectures. Posterior sampling is carried out using normalizing flows. Bounds are obtained via the two-stage procedure of \autoref{subsec:twostage}, which produces non-trivial results even for small training sets.}
  \label{fig:ub_robots}
\end{figure}

\paragraph*{\textbf{Evaluating the bounds}}
\autoref{fig:ub_robots} reports the test cost of $100$ controllers sampled from the posterior (circles) together with the corresponding upper bounds (black lines). Because the controllers in this experiment are high-dimensional, the bound from \autoref{corol:ub_qstar} cannot be computed via prior discretization; instead, we use the looser but tractable bounds from \autoref{corol:mcdim_ub}. The horizontal axis represents $\numRollouts$, and colors indicate whether the controller architecture is based on RENs or SSMs. To enable sampling a large number of controllers from the posterior, we approximate it using normalizing flows, since SVGD restricts sampling to at most $\numParticles$ different controllers, where $\numParticles$ denotes the number of particles.

In this high-dimensional setup, obtaining meaningful bounds is extremely challenging. Directly applying \autoref{corol:mcdim_ub} with the chosen prior yields bounds larger than $1$ for $\numRollouts \leq 128$. Since the cost is known to be bounded by $\costBound=1$, such results are trivial and uninformative. To address this, we employ the two-stage approach introduced in \autoref{subsec:twostage}, which uses part of the data to learn a data-dependent prior and then applies it in a second stage with the remaining data. The resulting bound, corresponding to the second stage, is expected to be tighter since the prior distribution is better aligned with the data. Indeed, as shown in \autoref{fig:ub_robots}, all of our computed bounds are below $1$, even with as few as $\numRollouts=4$ training sequences. Details on how the training samples are divided between the two stages are provided in \aref{app:optuna}.

In all cases, the true cost remains below the computed bound, confirming validity. Moreover, the bounds consistently tighten as $\numRollouts$ increases, demonstrating the benefits of larger training datasets. Finally, comparing architectures, we observe that SSM-based controllers both achieve lower cost and yield tighter bounds than RENs, highlighting their superiority in this setting.

\paragraph*{\textbf{Performance enhancement using bootstrapping}}
Since our framework produces a distribution over controllers, we can sample multiple candidates and select the best one to enhance performance. To preserve the validity of the bounds, note that \autoref{corol:ub_qstar} holds with probability $1-\confidence$ for a single controller sampled from $\Q^*$. If $\numQSamples \in \mathbb{N}$ controllers are sampled, the probability that the bounds hold for all of them is $1 - \numQSamples \confidence$, and thus also $1 - \numQSamples \confidence$ for the selected best controller. Consequently, by setting $\confidence^\prime = \confidence / \numQSamples$, sampling $\numQSamples$ controllers, and choosing the best one, the bounds continue to hold with probability $1-\confidence$ for the selected controller.

The next key question is how to select the best controller. While choosing the one with the lowest training cost is natural, it risks overfitting. Instead, we propose a bootstrap-based method. Bootstrapping is a resampling technique that repeatedly draws samples with replacement from the training dataset to estimate a statistic and its confidence interval~\cite{James2013}. We use this approach to estimate the out-of-sample cost of each candidate controller, thereby guiding the selection toward the one expected to generalize best.

We empirically validate this approach using REN-based controllers with $\numRollouts = 32$, $\confidence = 0.1$, and the Normalizing Flows approximation method. Specifically, we sample $\numQSamples = 10$ controllers from the approximated posterior distribution corresponding to $\confidence^\prime = 0.1 / 10 = 0.01$. We then compare the best controller obtained through bootstrapping against $100$ controllers sampled directly from the posterior distribution with $\confidence = 0.1$ in terms of the test cost. The bootstrapping-based controller achieves an $8\%$ improvement over the average cost of the sampled controller and a $20\%$ improvement over the worst sampled controller. This demonstrates that bootstrapping robustifies our approach by avoiding poor-performing controllers, even though the probability of encountering such cases is low.

\paragraph*{\textbf{Scalability}}\looseness-1
Finally, we train a PAC-SNOC REN-based controller with approximately $10^4$ parameters using SVGD approximation with a single particle for $1000$ epochs. The resulting test cost improves by $7\%$ compared to the results in Table~\ref{tab:robots}, and the training time is $2583 \pm 330$ seconds, demonstrating that our framework effectively scales to large controllers.


\section{Conclusions}
\label{sec:conclusions}
This paper introduced a PAC-Bayesian framework for SNOC that provides generalization guarantees by relating the true cost of a controller to its empirical performance on finite datasets.
On the algorithmic side, we developed a principled methodology for learning control policies by sampling from an optimal posterior distribution, ensuring closed-loop stability through neural architectures with built-in guarantees. The framework naturally incorporates prior knowledge via the prior distribution and remains computationally feasible for complex tasks thanks to modern approximate inference techniques.
Experiments on synthetic and robotic tasks confirmed these benefits, showing improved generalization. Future work includes extending the analysis to infinite-horizon settings, studying dataset-size dependence more rigorously, and leveraging existing controllers for prior design.
\bibliographystyle{IEEEtran}
\bibliography{ref.bib}
\appendix
\subsection{Related work}

\subsubsection{Learning stabilizing controllers}\label{app:relatedworkPB}
In general, stability can be imposed in an SNOC problem~\eqref{eq:SNOC} in two ways.
The first one is to resort to 
constrained optimization approaches that ensure global or local stability by enforcing appropriate Lyapunov-like inequalities during optimization~\cite{berkenkamp2018safe,gu2021recurrent,pauli2021offset}.
However, enforcing such constraints often becomes a computational bottleneck in complex applications. 
This issue has been highlighted in~\cite{dawson2023safe}, discussing that popular stability verification methods are feasible for controllers with an order of $10^2$ parameters.
The second possibility is to use
unconstrained optimization approaches searching within classes of control policies with built-in stability guarantees~\cite{wang2021learning,furieri2022distributed, NeurSLS, furieri2024learning, outputSLS}.
These methods allow learning controllers through optimization algorithms based on gradient descent, without imposing further constraints to ensure stability.
Additionally, unconstrained approaches usually have a much lower computational burden than constrained ones~\cite{NeurSLS, furieri2024learning, outputSLS}.
In \autoref{sec:approx}, we show that unconstrained approaches also enable the use of algorithms for sampling the posterior, such as SVGD.
Moreover, in \autoref{subsec:experiments_robots}, we demonstrate the scalability of these approaches by training a stabilizing controller with more than $10^4$ parameters.

\subsubsection{PAC-Bayesian control}\label{app:relatedworkPAC}
PAC-Bayesian bounds have not been widely applied to SNOC problems, with the exception of~\cite{majumdar2021pac}, which studied them in a different context, focusing on zero-shot generalization across new environments. Other related works employ PAC-Bayesian analysis for learning control policies in Markov Decision Processes~\cite{J1,J2}.
However, these approaches typically assume finite action or state spaces, which are not well-suited for SNOC.

A key reason for the limited adoption of classical PAC-Bayesian bounds in control is that they integrate actions from all possible controllers under the posterior, which is computationally prohibitive and impractical, as discussed in \autoref{sec:introduction}.
We address this challenge by extending PAC-Bayesian bounds for randomized predictors to SNOC, enabling tractable analysis in settings with infinite and continuous state and action spaces.

\subsection{Mathematical proofs}
\subsubsection{Proof of \autoref{prop:twostage}}\label{app:proof_twostage}
By substituting the expression for $\P_2$ obtained in the first stage into the formula for $\Q_2$ in the second stage of \autoref{prop:twostage}, we have:
\begin{align}
    \Q_2(\contParam)
    =&
    \P(\contParam)
    \,
    \frac{
    e^{-{\tempGibbs_1} \costEmp(\contParam, \S_1)}
    e^{-{\tempGibbs_2} \costEmp(\contParam, \S_2)}
    }
    {
    Z_{\tempGibbs_1}(\P_1, \S_1)
    Z_{\tempGibbs_2}(\P_2, \S_2)
    }\cdot\label{eq:proof_prop2_expanded}
\end{align}

Next, we analyze the numerator. When setting $\tempGibbs_1 = \frac{\numRollouts_1}{\numRollouts} \, \tempGibbs$ and $\tempGibbs_2 = \frac{\numRollouts_2}{\numRollouts} \, \tempGibbs$ according to \autoref{prop:twostage}, the product in the numerator of \eqref{eq:proof_prop2_expanded} simplifies to:
\begin{align}
    e^{-{\tempGibbs_1} \costEmp(\contParam, \S_1)}
    &e^{-{\tempGibbs_2} \costEmp(\contParam, \S_2)}
    \nonumber\\
    &=
    e^{-{\frac{\numRollouts_1}{\numRollouts} \tempGibbs} \costEmp(\contParam, \S_1)
    -
    {\frac{\numRollouts_2}{\numRollouts} \tempGibbs} \costEmp(\contParam, \S_2)}
    \nonumber\\
    &=
    e^{-\frac{\tempGibbs}{\numRollouts}
    \bigl(
    \sum_{\noise_{T:0} \in \S_1} \costTraj(\contParam, \noise_{T:0})
    +
    \sum_{\noise_{T:0} \in \S_2} \costTraj(\contParam, \noise_{T:0})
    \bigr)
    }
    \nonumber\\
    &=
    e^{-\frac{\tempGibbs}{\numRollouts}
    \sum_{\noise_{T:0} \in \S_1 \cup \S_2} \costTraj(\contParam, \noise_{T:0})}
    = e^{-{\tempGibbs} \costEmp(\contParam, \S)},
    \label{eq:proof_prop2_merged}
\end{align}
where the first and last steps follow from the definition of $\costEmp$ in \eqref{eq:emp_cost}.

We now simplify the denominator:
\begin{align}
    Z_{\tempGibbs_1}(\P_1, &\S_1)
    Z_{\tempGibbs_2}(\P_2, \S_2)
    \nonumber\\
    =&
    Z_{\tempGibbs_1}(\P_1, \S_1)
    \E_{\contParam \sim \P_2}
    e^{-\tempGibbs_2 \costEmp(\contParam, \S_2)}
    \nonumber\\
    =&
    Z_{\tempGibbs_1}(\P_1, \S_1)
    \E_{\contParam \sim \P}
    \Bigl[
    \frac{
    e^{-{\tempGibbs_1} \costEmp(\contParam, \S_1)}
    }
    {
    Z_{\tempGibbs_1}(\P_1, \S_1)
    }
    e^{-\tempGibbs_2 \costEmp(\contParam, \S_2)}
    \Bigr]
    \nonumber\\
    =&
    \E_{\contParam \sim \P}
    e^{-{\tempGibbs_1} \costEmp(\contParam, \S_1)}
    e^{-\tempGibbs_2 \costEmp(\contParam, \S_2)}
    =
    \E_{\contParam \sim \P}
    e^{-{\tempGibbs} \costEmp(\contParam, \S)},
    \label{eq:proof_prop2_denom}
\end{align}
where the first equality uses the definition of the partition function in \eqref{eq:Z}, the second substitutes the expressions for $\P_1$ and $\P_2$, and the last step follows from \eqref{eq:proof_prop2_merged}.

Finally, by substituting \eqref{eq:proof_prop2_merged} and \eqref{eq:proof_prop2_denom} into \eqref{eq:proof_prop2_expanded}, we recover the desired result:
\begin{align*}
\Q_2(\contParam)
=
\P(\contParam)
\,
\frac{
e^{-{\tempGibbs} \costEmp(\contParam, \S)}
}
{
Z_{\tempGibbs}(\P, \S)
}
=
\Q^*(\contParam).
\end{align*}



\subsubsection{Proof of \autoref{corol:mcdim_ub}}\label{app:proof_mcdim_ub}
We first recall McDiarmid's inequality, which is used in our proof.
\begin{lemma}[McDiarmid's inequality~\cite{concentration}]\label{lemma:mcdim}
    Let $X_1$, $X_2$, $\dots$, $X_N$ be independent random variables taking values in a set $\mathcal{X}$, and let $\varphi : \mathcal{X}^N \rightarrow \mathbb{R}$ be a function satisfying the bounded difference condition for each $n \in \{1, \cdots, N\}$:
    \begin{align*}
        \bigl\vert \varphi(x_1, \dots, x_n, \dots, x_N) - \varphi(x_1, \dots, x_n', \dots, x_N) \bigr \vert \leq c_n ,
    \end{align*}
    for all $x_1, \dots, x_N, x_n' \in \mathcal{X}$.
    Then, for any $\epsilon > 0$,
    \begin{multline*}
    \Pr \Bigl[ \varphi(X_1, \dots, X_N) - \mathbb{E} \bigl[ \varphi(X_1, \dots, X_N) \bigr] \leq \epsilon \Bigr] 
    \\
    \geq 1- e^\frac{-2\epsilon^2}{\sum_{n=1}^N c_n^2}.
    \end{multline*}
\end{lemma}

Define $\varphi(\contParam_1, \cdots, \contParam_{\numPSamples}) = \ln \bigl( \hat{Z}_\tempGibbs(\contParam_1, \cdots, \contParam_{\numPSamples}, \S) \bigr)$, where $\hat{Z}_\tempGibbs$ is given in \eqref{eq:Zhat}.  
We first relate the expected value of $\varphi$ to $Z_\tempGibbs(\P, \S)$:
\begin{multline}
    \E_{\contParam_1, \cdots, \contParam_{\numPSamples} \sim \P^{\numPSamples}} \ln \bigl( \hat{Z}_\tempGibbs(\contParam_1, \cdots, \contParam_{\numPSamples}, \S) \bigr)
    \\
    \leq
    \ln \bigl(\E_{\contParam_1, \cdots, \contParam_{\numPSamples} \sim \P^{\numPSamples}}  \hat{Z}_\tempGibbs(\contParam_1, \cdots, \contParam_{\numPSamples}, \S) \bigr)
    \\
    = 
    \ln \bigl( Z_\tempGibbs(\P, \S) \bigr), \label{eq:boundEf}
\end{multline}
where the inequality follows from Jensen’s inequality~\cite{concentration}, and the equality from the fact that $\E[\hat{Z}_\tempGibbs] = Z_\tempGibbs$.

Next, given \autoref{assumption}, the function $\varphi$ satisfies the bounded difference condition of Lemma~\ref{lemma:mcdim} with $c_{\counterPSamples} = \ln \bigl( 1+ \frac{e^{\tempGibbs \costBound} - 1}{\numPSamples} \bigr)$ for all $\counterPSamples \in \{ 1, \cdots, \numPSamples \}$. Let $\epsilon = \ln \bigl( 1+ (e^{\tempGibbs \costBound} - 1)/\numPSamples \bigr) \sqrt{0.5 \, \numPSamples \ln(1/\confidence)}$, where $\confidence$ is the confidence level in \autoref{theo:pac}. Then, we obtain from \eqref{eq:boundEf} and McDiarmid's inequality (Lemma~\ref{lemma:mcdim}) that:
\begin{multline}
    Pr \Bigl[
        \ln \bigl( \hat{Z}_\tempGibbs(\contParam_1, \cdots, \contParam_{\numPSamples}, \S) \bigr) - \ln \bigl( Z_\tempGibbs(\P, \S) \bigr) \geq 
        \\
        \ln \bigl( 1+ \frac{e^{\tempGibbs \costBound} - 1}{\numPSamples} \bigr) \sqrt{\frac{\numPSamples}{2} \ln(\frac{1}{\confidence})}
    \Bigr]
    \leq
    \delta. \label{eq:proofCorol3}
\end{multline}
Finally, we combine this result with \autoref{corol:ub_qstar}.
Recall that the inequalities in \autoref{corol:ub_qstar} hold individually with probability at least $1-\confidence$ and jointly with probability at least $1-2\confidence$ over simultaneously sampling $\S \sim \D_{T:0}^\numRollouts$ and $\contParam \sim \Q^*$.
Since the probability events in \autoref{corol:ub_qstar} and \eqref{eq:proofCorol3} are defined over independent variables, we can invoke the union bound to combine them, which yields the desired result.%
\footnote{The union bound states that if $\varphi_1(x_1) \geq c_1$ with probability at least $1-\delta_1$ and $\varphi_2(x_2) \geq c_2$ with probability at least $1-\delta_2$, where $x_1$ and $x_2$ are independent, then both inequalities hold simultaneously with probability at least $1-\delta_1-\delta_2$.}

\subsection{Implementation details}\label{app:implementation}
\subsubsection{REN}
In~\cite{NeurSLS, furieri2024learning}, the mapping $\Emme_{\infty:0}^{\contParam}$ is implemented as a Recurrent Equilibrium Network (REN)~\cite{revay2023recurrent}, defined at each $t$ as:
\begin{equation}\label{eq:REN}
    \Emme_t^{\contParam} : \qquad
    \begin{bmatrix}
    \xi_{t+1} \\
    \zeta_t \\
    \inp_t
    \end{bmatrix} 
	=
    {\Omega(\contParam)}
    \begin{bmatrix}
    \xi_{t} \\
    \sigma(\zeta_t) \\ 
    \hat{\noise}_{t} 
    \end{bmatrix}
    \, ,
\end{equation}
where $\xi_t \in \R^{N\xi}$ is an internal state initialized at zero, and $\zeta_t \in \R^{N_\zeta}$ is an internal variable. The activation function $\sigma:\R \rightarrow \R$ is applied elementwise and must be piecewise differentiable with first derivatives in $[0, 1]$. The mapping $\Omega: \R^{\contParamDim} \rightarrow \R^{(N_\xi + N_\zeta + \inpDim) \times (N_\xi + N_\zeta + \stateDim)}$, provided in~\cite{furieri2024learning}, ensures that $\zeta_t$ is uniquely defined at every $t \in \N_0$, and that~\eqref{eq:REN} is $\ell_2$-stable for all $\contParam \in \R^\contParamDim$, i.e., $\contParamSet = \R^\contParamDim$. Together, \eqref{eq:neurSLS} and \eqref{eq:REN} specify a dynamical neural controller that guarantees closed-loop stability for any $\contParam \in \R^\contParamDim$.


In \autoref{subsec:experiments_robots}, we use RENs with $(N_\xi, N_\zeta) = (8, 8)$, resulting in $\contParamDim = 864$ parameters. For the scalability test, we increase to $(N_\xi, N_\zeta) = (32, 32)$, yielding $\contParamDim = 11040$ parameters.

For the empirical approach, the REN parameters are initialized by sampling from a Gaussian distribution with a zero mean and a diagonal covariance matrix $\sigma^2 I$. The variance $\sigma^2$ is a tunable hyperparameter, as detailed in \aref{app:optuna}. In contrast, when training with PAC-SNOC, explicit initialization is not required since the REN parameters are sampled directly from the initial distribution over controllers.

\subsubsection{SSM}
As discussed in \autoref{subsec:controller_parametrization}, the only 
requirement on $\Emme_{\infty:0}^{\contParam}$ is that it is $\ell_p$-stable. 
This requirement can be satisfied by a recently proposed class of Structured 
State-Space Models (SSMs), namely the L2RU architecture~\cite{massai2025free}. 

The L2RU architecture consists of a sequence of \emph{state-space layers}, where each layer is a sequence of an $\ell_2$-bounded discrete-time LTI system with a Lipschitz-bounded nonlinearity. Linear pre- and post-processing modules are also applied to the input and output. Crucially, the L2RU parametrization ensures a prescribed input-output $\ell_2$-bound for any choice of parameters across all modules. This design guarantees input-output stability without the requirement of satisfying explicit constraints during optimization, so that one has $\contParamSet = \R^\contParamDim$. 

In \autoref{subsec:experiments_robots}, we instantiate the controller with two state-space layers, each with a hidden state of dimension $8$. The nonlinearities are implemented as Coupling Layers~\cite{dinh2017density, Papamakarios21Normalizing} with hidden layers of dimension $8$. This configuration results in a dynamical neural controller with $\contParamDim = 896$ parameters that guarantees closed-loop stability for any $\contParam \in \R^\contParamDim$.

When training with the empirical approach, the parameters of SSMs are initialized following the procedure in~\cite{gu2022efficiently, massai2025free}, where the eigenvalues of each state-space layer are sampled uniformly at random inside a disk centered at the origin with outer radius $1$ and a non-zero inner radius. This prevents eigenvalues with small modulus, which may induce oscillations in the SSM output~\cite{gu2022efficiently}. The inner radius is treated as a tunable hyperparameter, as discussed in \aref{app:optuna}. Similar to RENs, explicit initialization is not required when training with PAC-SNOC since the SSM parameters are sampled directly from the initial distribution over controllers.



\subsubsection{SVGD}\label{app:svgd}
As discussed in \autoref{sec:approx}, Stein Variational Gradient Descent (SVGD)~\cite{liu2016stein} approximates the optimal posterior distribution $\Q^*$ as a discrete distribution over a set of particles $\{\particle_i\}_{i=1}^{\numParticles}$. Each particle $\particle_i \in \contParamSet$ represents the parameters of a controller and is initialized by sampling from the prior distribution $\P$. 
At each iteration, the particles are updated according to:
\begin{align*}
    \particle_i \leftarrow& \particle_i + \eta \, \varphi(\particle_i), 
    \\
    \varphi(\particle) =& \frac{1}{\numParticles} \sum_{j=1}^\numParticles
    \Big[ \kappa(\particle_j,\particle) \nabla_{\particle_j} \log \Q^*(\particle_j) 
    + \nabla_{\particle_j} \kappa(\particle_j,\particle) \Big],
\end{align*}
where $\eta>0$ is the learning rate and $\kappa(\cdot,\cdot)$ is a positive definite kernel that encourages diversity among the particles by repelling them from one another.
We employ the RBF kernel with bandwidth selected using the median heuristic~\cite{liu2016stein}. 

Importantly, the gradient $\nabla_{\particle} \log \Q^*(\particle)$ requires only the unnormalized density of $\Q^*$ in~\eqref{eq:qstar}, thus avoiding computation of the partition function $Z_\tempGibbs$. After convergence, the approximate posterior is given by the uniform distribution over the $\numParticles$ particles. The learning rate is treated as a hyperparameter, tuned as described in \aref{app:optuna}.

\subsubsection{Normalizing flows}\label{app:nf}
Normalizing flows~\cite{rezende2015variational} provide an alternative to SVGD for approximating the optimal posterior $\Q^*$ by learning a parametric distribution with tractable density evaluation and sampling. The key idea is to begin from a simple base distribution $p_0(z)$ over $z \in \R^d$, typically a multivariate Gaussian, and map it to a complex distribution through a sequence of invertible transformations. Formally, letting $z_0 \sim p_0(z)$ and $z_L = f_L \circ \cdots \circ f_1(z_0)$, the resulting density is obtained via the change-of-variables formula:
\begin{equation*}
    \log q(z_L) = \log p_0(z_0) - \sum_{\ell=1}^L \log \Big| \det \nabla f_\ell(z_{\ell-1}) \Big|.
\end{equation*}

In our framework, the transformed variable $z_L$ corresponds to the controller parameters, i.e., $z_L \equiv \contParam$. The flow parameters, along with the mean and covariance of the base distribution, are optimized by minimizing the KL divergence between the induced distribution $q$ and the target Gibbs posterior $\Q^*$. Training is performed via stochastic gradient descent, with the learning rate tuned as a hyperparameter following \aref{app:optuna}.

We use $L=16$ transformations and define each $f_\ell$ as a \emph{planar flow}~\cite{rezende2015variational},
$
    f(z) = z + u \, h(w^\top z + b),
$
where $u, w \in \R^d$, $b \in \R$, and $h$ is the hyperbolic tangent function. Planar flows offer a balance between flexibility and computation, as their Jacobian determinant is available in closed form.
We used the open-source NormFlows package~\cite{Stimper2023} in our code.

We observed that convergence is sensitive to the choice of base distribution. To address this, we adopt a data-driven initialization: we first train controllers using the empirical method on a single fixed data point with five different random seeds, then center the base distribution at the mean of the resulting parameters. The covariance is chosen as a diagonal matrix, where the diagonal entries are obtained by scaling the sample variance of these parameters by a constant factor greater than one. This scaling factor is treated as a hyperparameter and tuned as described in \aref{app:optuna}.

\subsection{Hyperparameter tuning}
\label{app:optuna}
In our framework, hyperparameters include the standard deviation of the Gaussian initialization for RENs, the inner radius of the eigenvalue initialization disk for SSMs, the scaling constant of the base distribution covariance for training normalizing flows, the standard deviation of the prior distribution for PAC-SNOC, and the learning rate for all methods. 

We tune these hyperparameters using the open-source \emph{Optuna} package~\cite{optuna}, which performs automated hyperparameter search via Bayesian optimization. The objective function is defined as the validation cost over a held-out subset of $25\%$ of training trajectories, and the best-performing hyperparameter configuration over $10$ trials is selected. For reproducibility, we specify the search ranges and the final selected hyperparameters in the codebase accompanying this paper.

When using the two-stage approach, the number of data points used in each stage can also be tuned. For this, we performed a grid search aimed at minimizing the constant terms in \autoref{corol:mcdim_ub}, since these constants are generally dominant. From the three values yielding the smallest constant terms, we then computed the full bound and selected the configuration with the tightest result. Note that computing the full bound requires two rounds of posterior inference, which makes a direct grid search over the bound in \autoref{corol:mcdim_ub} impractical. This motivates our procedure, where the grid search is first performed over the constant term only, and then the best candidates are fully evaluated. In principle, Bayesian optimization could also be used for this purpose, though we did not explore this option.

\begin{IEEEbiography}[{\includegraphics[width=1in,height=1.25in,clip,keepaspectratio]{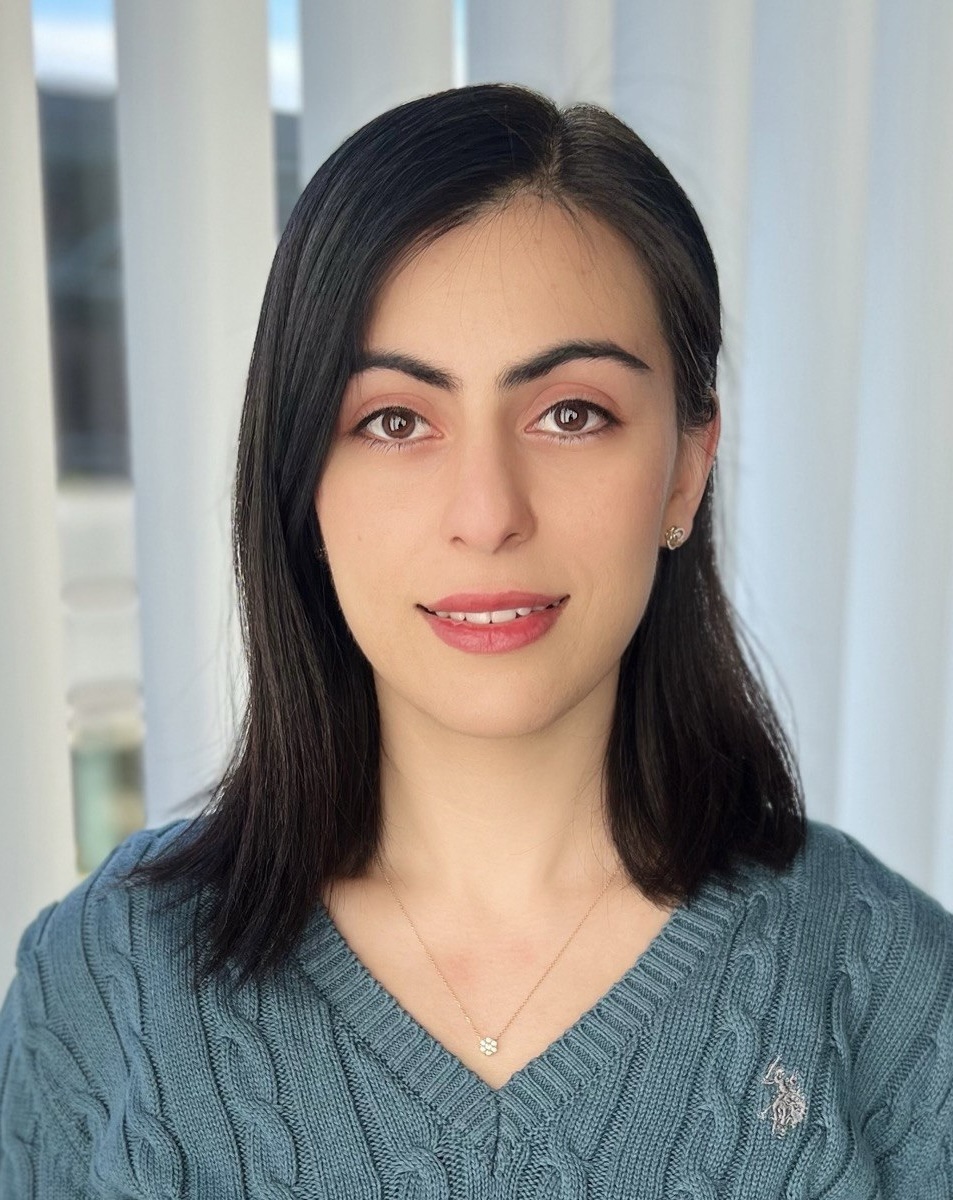}}]{Mahrokh Ghoddousi Boroujeni}
is a Ph.D. candidate in the Dependable Control and Decision group (DECODE) at École Polytechnique Fédérale de Lausanne (EPFL), Lausanne, Switzerland. She is also a member of the National Centre of Competence in Research (NCCR) Automation.
She received her B.Sc. degree in Electrical Engineering and Computer Science from Sharif University of Technology, Tehran, Iran, in 2020. Her research interests include learning-based control, stability and generalization guarantees, probabilistic modeling, and meta-learning.
\end{IEEEbiography}

\begin{IEEEbiography}[{\includegraphics[width=1in,height=1.25in,clip,keepaspectratio]{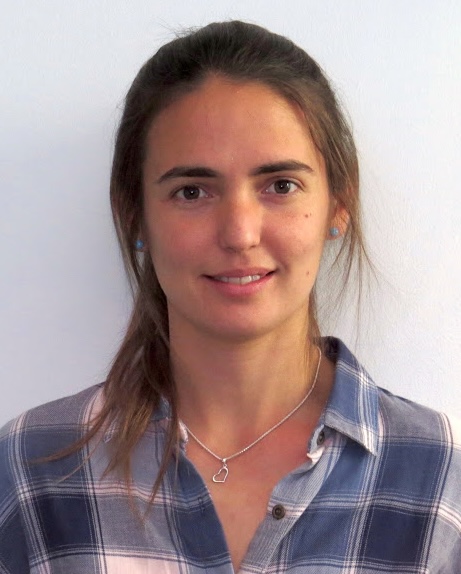}}]{Clara Lucía Galimberti}
received the Ph.D. degree from the École Polytechnique Fédéral de Lausanne, Lausanne, Switzerland, in 2024, and the degree in electronic engineering from the Universidad Nacional de Rosario, Rosario, Argentina, in 2018.
Since March 2025, she has been a researcher at Dalle Molle Institute for Artificial Intelligence (IDSIA USI-SUPSI), University of Applied Sciences and Arts of Southern Switzerland, Lugano, Switzerland.
Her research interests include optimal and learning-based control, and machine learning.
\end{IEEEbiography}

\begin{IEEEbiography}[{\includegraphics[width=1in,height=1.25in,clip,keepaspectratio]{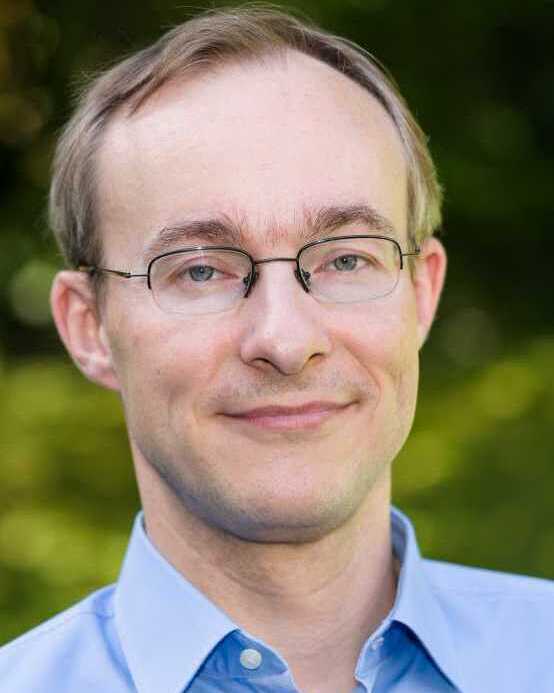}}]{Andreas Krause}
(Fellow, IEEE) Received a Diplom in computer science and mathematics from the Technical University of Munich, Munich, Germany, in 2004, and the Ph.D. degree in computer science from Carnegie Mellon University, Pittsburgh, PA, USA, in 2008. He is currently a Professor of computer science at ETH Zurich, Zurich, Switzerland, where he also serves as the Chair of the ETH AI Center. He was an Assistant Professor of computer science with Caltech, Pasadena, CA,
USA.
Dr. Krause is an ACM Fellow, IEEE Fellow, and ELLIS Fellow. His research on machine learning and adaptive systems has received multiple awards, including the ACM SIGKDD Test of Time award 2019 and the ICML Test of Time award 2020. He was the Program Co-Chair for ICML 2018, General Chair for ICML 2023, and Action Editor for the Journal of Machine Learning Research. In 2023-2024, he served on the United Nations’ High-level Advisory Body on AI.
\end{IEEEbiography}

\begin{IEEEbiography}[{\includegraphics[width=1in,height=1.25in,clip,keepaspectratio]{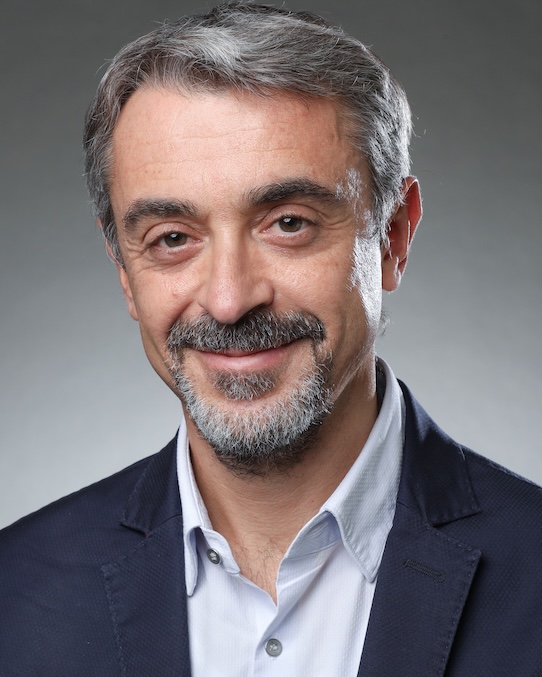}}]{Giancarlo Ferrari Trecate}
(SM'12) received a Ph.D. in Electronic and Computer Engineering from the Università Degli Studi di Pavia in 1999. Since September 2016, he has been a Professor at EPFL, Lausanne, Switzerland. In the spring of 1998, he was a Visiting Researcher at the Neural Computing Research Group, University of Birmingham, UK. In the fall of 1998, he joined the Automatic Control Laboratory, ETH, Zurich, Switzerland, as a Postdoctoral Fellow. He was appointed Oberassistent at ETH in 2000. In 2002, he joined INRIA, Rocquencourt, France, as a Research Fellow. From March to October 2005, he was a researcher at the Politecnico di Milano, Italy. From 2005 to August 2016, he was Associate Professor at the Dipartimento di Ingegneria Industriale e dell'Informazione of the Università degli Studi di Pavia.
His research interests include scalable control, machine learning, microgrids, networked control, and hybrid systems.
Giancarlo Ferrari Trecate is the founder and current chair of the Swiss chapter of the IEEE Control Systems Society. He is Senior Editor of the IEEE Transactions on Control Systems Technology and has served on the editorial boards of Automatica and Nonlinear Analysis: Hybrid Systems. 
\end{IEEEbiography}

\vfill

\end{document}